\documentclass{llncs}

\usepackage{amsmath,amssymb,amsfonts}
\usepackage[dvipdfmx]{graphicx}
\usepackage{bm}
\usepackage{listings}
\usepackage{xcolor}
\usepackage{url}
\usepackage[caption=false]{subfig}
\usepackage{multirow}
\usepackage[linesnumbered,ruled,vlined]{algorithm2e}
\usepackage{mathtools}
\usepackage{wrapfig}
\usepackage{here}
\usepackage{intmacros}

\usepackage{article}

\begin{document}

\mainmatter

\title{Approximate Translation from Floating-Point to Real-Interval Arithmetic}

\author{Daisuke Ishii\inst{1} \and Takashi Tomita\inst{1} \and Toshiaki Aoki\inst{1}}
\institute{Japan Advanced Institute of Science and Technology\\
  \email{\{dsksh,tomita,toshiaki\}@jaist.ac.jp} }

\maketitle

\begin{abstract}
    Floating-point arithmetic (FPA) is a mechanical representation of real arithmetic (RA), where each operation is replaced with a rounded counterpart.
    Various numerical properties can be verified by using SMT solvers that support the logic of FPA. 
    However, the scalability of the solving process
    remains limited when compared to RA.
    %
    %
    In this paper, we present a decision procedure for FPA that takes advantage of the efficiency of RA solving.
    The proposed method abstracts FP numbers as rational intervals and FPA expressions as interval arithmetic (IA) expressions;
    then, we solve IA formulas to check the satisfiability of an FPA formula using an off-the-shelf RA solver (we use \textsc{CVC4} and \textsc{Z3}).
    In exchange for the efficiency gained by abstraction, 
    the solving process becomes quasi-complete; 
    we allow to output $\Unknown$ when the satisfiability is affected by possible numerical errors.
    Furthermore, our IA is meticulously formalized to handle the special value $\NaN$.
    We implemented the proposed method and compared it to four existing SMT solvers in the experiments.
    As a result, we confirmed that our solver was efficient for instances where rounding modes were parameterized.
\end{abstract}

\section{Introduction}

A key technique to perform calculations on reals efficiently is 
\emph{floating-point arithmetic} (FPA; Sect.~\ref{s:fpa})~\cite{IEEE754,Muller2010},
although there will be numerical errors caused by rounding reals into FP numbers.
The decision procedure on the logical theory of FPA is important for verifying numerical programs, hardware models, etc., while accounting numerical errors.
Indeed, such a theory and dedicated decision procedures have been developed and implemented in recent SMT solvers (e.g. \cite{Brain2019}).
Many solvers are based on a technique called \emph{bit blasting}~\cite{Clarke2004,Brillout2009} that encodes a satisfiability problem on FPA into that on bit vectors (BVs).
Despite the high performance of SAT solvers and several improvements (Sect.~\ref{s:related}),
the FPA solvers are less scalable than the real arithmetic (RA) solvers; especially when solving instances described by the same arithmetic formulas, the former is slower (sometimes in orders of magnitude) than the latter.

This paper aims to realize an efficient method by using RA solvers instead of bit blasting.
The proposed method represents an FPA expression with a rational interval that encloses every valuation in FP numbers for the expression.
This abstraction, which assumes arbitrary rounding modes and mild estimation of rounding errors,
slightly limits the method's target problem and completeness.
However, we expect to solve practical FPA problems with this approach by leveraging the efficiency of an off-the-shelf RA solver.
In addition, it is interesting to compare bit blasting with our method, as it explores the optimal decision procedure at the boundary between Boolean and continuous domains.

The contributions described in this paper are as follows:
\begin{itemize}
    \item \emph{A method to solve FPA formulas by encoding them into formulas on real intervals}.
        We formalize interval arithmetic (IA) for this purpose that handles the special FP number $\NaN$ (not a number) correctly (Sect.~\ref{s:ia}).
        A linear function for error estimation, an \emph{interval extension} scheme for FPA formulas, and an encoding method from the FPA logic to the real-interval logic are presented.
        \emph{Weak} and \emph{strong} modes are used for encoding, and their correspondence with $\delta$-variants~\cite{Gao2012} is discussed.
        The method is implemented as a tool that translates between SMT-LIB descriptions for FPA formulas and for RA formulas that embed the interval extension (Sect.~\ref{s:impl}).
        We also implement a script that solves an FPA formula using Z3.
    %
    \item \emph{Experiments to confirm the efficiency of the proposed method by comparison with four other SMT solvers} (Sect.~\ref{s:xp}).
        We prepared FPA problem instances in three sets, in which a set is a typical FPA benchmark and two sets consist of instances with no rounding mode setting.
        In the experiments, we obtained promising results when comparing our method to existing FPA solvers.
        We confirmed that our solver solved the most number of problems for each set except for the FPA benchmark.
        We also confirmed that the number of inconclusive ($\Unknown$) results by our solver were small ($< 10\%$) in all but one of the six categories.
\end{itemize}

\subsubsection{Examples.}
Let $f$ be a real function.
Suppose we want to check the satisfiability of an FPA formula $\phi_\mathbb{F} := f_\mathbb{F} >_\mathbb{F} +0_\mathbb{F}$,
which is a direct translation of an RA formula $\phi := f > 0$,
obtained by replacing every syntactic element in $\phi$ with an FPA counterpart (with a rounding mode configuration).
We can feed them to an SMT solver equipped with FPA and RA solvers;
then, the solving process for $\phi_\mathbb{F}$ is often less efficient than solving $\phi$.
When we can estimate an error bound $\delta := |f-f_\mathbb{F}|$,
checking variant formulas $\phi^- := f > -\delta$ and $\phi^+ := f > \delta$ by the RA solver might be more efficient.
If $\phi^-$ is $\Unsat$ or $\phi^+$ is $\Sat$, then so is $\phi_\mathbb{F}$;
otherwise, this method could answer ``$\Unknown$'' ($f$ can be in the $\delta$-neighborhood of zero).
%
The proposed method in this paper translates $f$ into an expression $\f$ based on an IA, which evaluates to an interval that overapproximates $f_\mathbb{F}$ assuming any rounding modes;
then, $\delta$ is obtained as the width of $\f$.

The case $f_\mathbb{F}$ evaluates to the special value $\NaN$ makes this method complicated.
The satisfiability of a negative predicate $\phi'_\mathbb{F} := \neg(f_\mathbb{F} \leq_\mathbb{F} +0_\mathbb{F})$
might be checked using the same formulas $\phi^-$ and $\phi^+$.
However, if we assume that $f_\mathbb{F}$ evaluates to either $\NaN$ or other FP numbers, it is not correct;
$\phi'_\mathbb{F}$ is $\Sat$ regardless of other assignments, because ${\NaN \leq_\mathbb{F} +0_\mathbb{F}}$ does not hold.
Therefore, we use variant formulas
$\phi'^-$ and $\phi'^+$,
prepared specifically for negative predicates.
When $f_\mathbb{F}$ \emph{can be} $\NaN$, 
$\phi'^-$ holds regardless of other assignments so that the satisfiability by a $\NaN$ assignment is taken into account
(satisfiability of $\phi'^+$ depends on the other assignments).

%



%

\section{Related Work}
\label{s:related}

The FPA theory solvers contained in SMT solvers have been actively developed over the last ten years or so, as summarized in \cite{Brain2019,Zitoun2020}.

\emph{Bit blasting}~\cite{Clarke2004,Brillout2009} is a major approach applied in many SMT solvers including \textsc{Z3}, \textsc{CVC4} and \textsc{MathSAT}.
It converts an FPA formula to a Boolean formula by encoding an FP number into a set of Boolean variables and FP operator circuits into Boolean formulas.
Because the size of an encoded formula easily becomes large, many approximation methods have been studied (e.g. \cite{Brillout2009,Haller2012,Brain2014,Scheibler2016,Ramachandran2016}).
Brain et al.~\cite{Brain2019} have implemented a reference bit-blasting engine included in \textsc{CVC4}.
This paper proposes a non-bit-blasting solver based on an RA solver, with competing results in the experiments.
The number of encoded real variables is proportional to the number of original FP variables, and we confirmed that memory usage is lower on average than in other solvers.

There are several works that encode FPA in RA.
Leeser et al.~\cite{Leeser2014} have proposed precise FPA embedding in an extended RA.
The performance of their solver \textsc{Realizer} 
was not competitive 
in our preliminary trial.
A mixed-real-FPA~\cite{Ramachandran2016,Salvia2019} have been proposed to encode FPA, where some formulas are approximated by real formulas with rounding operations removed and other formulas are left unchanged.
Their procedure tries to simplify the formula (and then solve it) by searching for such a formula that retains the same solution as the original.
Zelj\'{i}c et al.~\cite{Zeljic2018} have proposed an approximation framework based on a similar idea.
They examined the mixed-real-FPA as an approximation domain.
In contrast to \cite{Ramachandran2016,Zeljic2018,Salvia2019}, our method encodes an overapproximation of FPA formulas in RA; the result of solving the encoded formula is sound, whereas the above methods require verification after solving.

In decision procedures, IA-based techniques~\cite{Moore1966,Tucker2011} play a crucial role in various ways.
For RA logic formulas, 
there are solvers based on intervals bounded by FP numbers~\cite{Older1993,Michel2001,Franzle2007,Gao2012IJCAR,Tung2017};
in contrast, we approximate FPA formulas using real intervals.
IA-based decision procedures tend to be incomplete but can be \emph{$\delta$-complete}~\cite{Gao2012}; the same idea is applied in our encoding method.

IA is used frequently in FPA solvers to accelerate their process by approximating FP numbers.
Typically, it is coupled with bit blasting and algorithms such as CEGAR~\cite{Brillout2009} and non-chronological backtracking~\cite{Haller2012,Brain2014,Scheibler2016}.
\textsc{MathSAT} implements the method in \cite{Brain2014}.
\textsc{ObjCP}~\cite{Zitoun2017,Zitoun2020} and \textsc{Colibri}~\cite{Marre2017} are CP-based solvers implementing constraint propagation algorithms and other techniques e.g. diversification~\cite{Zitoun2020} and distance constraints~\cite{Marre2017}.
Bit-blasting solvers and CP solvers use intervals bounded by FP numbers, whereas ours uses intervals bounded by rational numbers.
\textsc{Colibri} also uses integer and real intervals, but the details have not been made public.

IA is also used in the static analysis of numerical programs~\cite{Daumas2004,Goubault2006,Solovyev2018,Titolo2018,Darulova2018}.
It is typically used for abstraction of numerical computation and to compute bounds for rounding errors.
Computation of tight bounds using Taylor expansion~\cite{Solovyev2018} and Affine arithmetic~\cite{Darulova2018} have been proposed;
our method can adopt these methods to improve the accuracy.
In terms of abstraction of FP expressions, Sect.~\ref{s:ia} can be regarded as a variant of the formalization in e.g. \cite{Titolo2018}.
However, our method differs in that 1) we aim at efficient solving of FPA logic formulas and 2) we formalizes NaN cases that is essential in the FPA logic.

Another branch of solvers applies an approach that encodes an axiomatization of FP numbers in the theory of reals and integers~\cite{Daumas2004,Brain2015b,Conchon2017}.
Our method can be considered to be in line with this approach, except that ours axiomatizes an interval extension of FPA in which rounding operations are overapproximated.


%

\section{Floating-Point Arithmetic}
\label{s:fpa}


{FP numbers}~\cite{IEEE754,Muller2010} are machine-representable approximations of real numbers.
They are represented as BVs, and we consider various sets of FP numbers parameterized with the size of BVs (we limit the radix to $2$). 
%
\begin{definition}
    Let $\mathit{eb}$ and $\mathit{sb}$ be the sizes of \emph{exponent} and \emph{significand} bits, respectively.
    An \emph{FP number} is represented by a pair $(M,e)$ of two integers such that $|M| \leq 2^{\mathit{sb}}-1$ and $e \in [1\!-\!e_\Max,e_\Max]$, where $e_\Max = 2^{\mathit{eb}-1}-1$;
    it is interpreted as a real number $M \!\times\! 2^{e-\mathit{sb}+1}$.
    In addition, we use special data.
    There are two \emph{signed zeros} $-0$ and $+0$ (we denote either of them by 0 if the difference does not matter). 
    \emph{Infinities} $-\infty$ and $+\infty$ represent numbers outside the representable bounds.
    Another special value $\NaN$ represents the result of exceptional evaluations.
    $\mathbb{F}_{\mathit{eb},\mathit{sb}}$ and $\mathbb{F}_{\mathit{eb},\mathit{sb}}^*$ denote the sets of FP numbers with fixed bit sizes, where $\mathbb{F}_{\mathit{eb},\mathit{sb}} = \mathbb{F}_{\mathit{eb},\mathit{sb}}^* \setminus \{-\infty,+\infty,\NaN\}$.
    We simply denote $\mathbb{F}$ and $\mathbb{F}^*$ when bit sizes are not important.
\end{definition}

For 64-bit double-precision FP numbers, $\mathit{eb}=11$ and $\mathit{sb}=53$.
In a multi-sort context, we also denote an FP number $n$ by $n_\mathbb{F}$.

\begin{definition} \label{d:v}
    We consider the sets of extended reals $\mathbb{R}^+ := \mathbb{R}\cup\{-\infty,+\infty\}$ and $\mathbb{R}^* := \mathbb{R}^+ \cup \{\NaN\}$.
    We interpret an FP number by mapping to the corresponding element in $\mathbb{R}^*$ using the function $v : \mathbb{F}^* \to \mathbb{R}^*$.
    $v(\mp0)$ evaluates to $0$.
\end{definition}


\begin{wrapfigure}[9]{r}{0.4\textwidth}
\centering
    \vspace{-3em}
    \includegraphics[width=.35\textwidth]{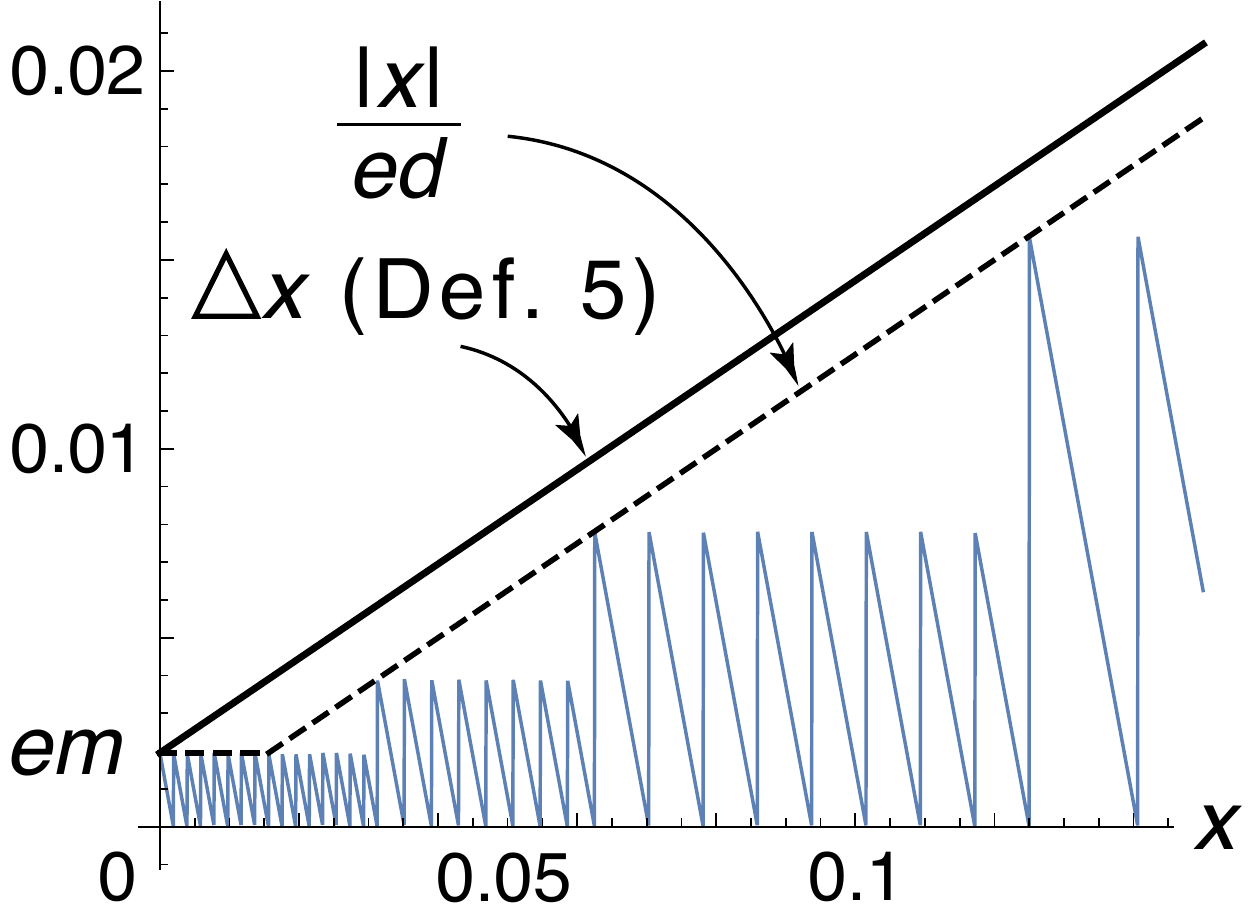} 
    \vspace{-.5em}
    \caption{\label{f:err} Rounding error.}
\end{wrapfigure}

Distribution of FP numbers is not uniform, and errors increase as the value increases.
Fig.~\ref{f:err} illustrates the errors when $x$ is rounded in $\mathbb{F}_{4,4}$.
Errors can be estimated using parameters $\mathit{ed} := 2^{\mathit{sb}-1}$, which is the inverse of the slope that approximates the error function,
and $\mathit{em} := {(\mathit{ed}-1)}/{(2^{e_\Max-1} \mathit{ed})}$, which is the error bound for subnormal FP numbers such that $|M| < 2^{\mathit{sb}-1}$.

In FPA (arithmetic with FP numbers), we apply four operators $\circ \in \{+,-,\times,\AB\div\}$, other operators e.g. absolute value $|\!\cdot\!|$, and comparison operators.
Although their semantics follows from the real interpretation, FP numbers are not closed under those operations, so the results are \emph{rounded} to neighboring FP numbers, causing numerical errors.
Each rounded operation should be associated with one of the six \emph{rounding modes}, 
e.g., the mode 
``\emph{round to nearest} \emph{ties to even (RNE)}'' 
rounds to a closest FP number; when two FP numbers are of the same distance, one with an even $M$ is chosen.
The set of modes are denoted by $\mathbb{M}$.
In the SMT-LIB format, there are two equation operators: \lstinline{fp.eq} and \lstinline{=}; in the sequel, we denote them by $=$ and $\equiv$, respectively. The main difference between the two is that $\NaN = \NaN$ does not hold but $\NaN \equiv \NaN$ does, and the former does not distinguish between the zeros but the latter does.

The semantics of FPA is specified in the IEEE-754 standard~\cite{IEEE754}, formalized in several works e.g. \cite{Brain2015b,Conchon2017}, and mechanically specified by the SMT-LIB \verb|FloatingPoint| theory.\footnote{\url{https://smtlib.cs.uiowa.edu/theories-FloatingPoint.shtml}.} 
Notably, for the special data, dedicated arithmetic rules are applied, e.g.,
$+\infty - +\infty$ and $+\infty\times0$ evaluate to $\NaN$,
and
$x = \NaN$ does not hold, where $x$ is an arbitrary FP number.

We consider logic formulas involving FPA predicates.
\begin{definition} \label{d:fpa}
    The grammar of \emph{FPA formulas} is as follows:
    \begin{align*}
        F ~&::=~ T \Rel T 
        ~|~ \neg F ~|~ F \lor F \\
        T ~&::=~ c ~|~ \mathit{id} ~|~ \mathit{uop}(T) ~|~ \mathit{bop}(\mathit{rm}, T, T)
    \end{align*}
    where $\Rel \in \{\equiv, =,\geq,>\}$, 
    $c$ is a literal of a sort $\mathbb{F}_{\mathit{eb},\mathit{sb}}$, and
    $\mathit{id}$ is a variable name.
    $\mathit{uop}$ are unary operations $-(\cdot)$ and $|\!\cdot\!|$, which do not require rounding, and $\mathit{bop}$ are binary operations $+$, $-$, $\times$, $\div$, associated with a rounding mode.
\end{definition}
We assume that formulas are well-sorted, insisting that every subformula is sorted in $\mathbb{F}_{\mathit{eb},\mathit{sb}}$ with the same $\mathit{eb}$ and $\mathit{sb}$.
In the other sections, we also denote FPA constructs such as $\Rel$ and $c$ by $\Rel_\mathbb{F}$ and $c_\mathbb{F}$ to indicate the sort.

The SMT-LIB's FPA theory supports multi-precision FPA and multi-sort FPA mixed with BV, integers and reals, wherein additional sort-conversion operators are needed.
The theory also provides additional predicates, e.g. \lstinline{fp.isNaN} and \lstinline{fp.isNegative}.
Our implementation (Sect.~\ref{s:impl}) supports many of these features, but some are left for future work.

\section{Abstraction of FPA with Interval Arithmetic}
\label{s:ia}

Sect.~\ref{s:ia:basic} introduces basics about IA and defines an interval extension of FP operators.
In Sect.~\ref{s:ia:logic}, we consider logic formulas involving interval predicates and how to convert an FPA formula into that system.
Then, the soundness basis of the proposed method is described.

\subsection{Interval Arithmetic}
\label{s:ia:basic}

IA~\cite{Moore1966} is a traditional method for the abstraction of continuous domains e.g. reals.
In this paper, we apply it to approximate FPA.
We introduce intervals in the domain $\mathbb{R}^*$ that approximates the values handled in FPA (i.e. the codomain of $v$ in Def.~\ref{d:v}).

\begin{definition}
    \emph{Intervals} are
    $\x = [\LB{x},\UB{x}] := \{\tilde{x} \in \mathbb{R}^+ ~|~ \LB{x} \leq \tilde{x} \leq \UB{x}\}$,
    where $\LB{x}, \UB{x} \in \mathbb{Q}\cup\{-\infty,+\infty\}$, and $\LB{x} \leq \UB{x}$.
    \emph{Point} intervals such that $\LB{x}=\UB{x}=x$ are also denoted by $[x]$.
    Furthermore, 
    we consider intervals that additionally contain $\NaN$; they are denoted either as $[\LB{x},\UB{x}] \cup \{\NaN\}$ or as $\x$ when considering genric intervals.
    %
    We denote the set of intervals by $\mathbb{I}^*$.
\end{definition}
Interval $[-\infty,+\infty]$ represents the entire domain $\mathbb{R}^+$ 
bounded by $-\infty$ and $+\infty$.
Point interval $[+\infty]$ represents the set $\{+\infty\}$.
%
We do not consider the empty set and $\{\NaN\}$ as intervals to make the analysis simple in return for a slight increase of abstraction.

%
To abstract the rounding of a real number $\tilde{x} \in \mathbb{R}^+$ to an FP number $x \in \mathbb{F}^+_{\mathit{eb},\mathit{sb}}$ (with an arbitrary mode), we consider an interval $\x$ such that $v(x) \in \x$.
It is preferable to have a tight $\x$, but accurate encoding of its bounds will be costly when later handling with SMT solvers;
thus, we use a linear approximation of rounding operators, at the expense of inaccuracy.
They are based on the numerical error analysis in Sect.~\ref{s:fpa}.
\begin{definition} \label{d:rounding}
    We assume a set of FP numbers $x \in \mathbb{F}^+_{\mathit{eb},\mathit{sb}}$.
    The \emph{rounding operators} $\RndD x$ and $\RndU x$ are defined respectively by:
    \begin{align*}
        \RndD x &:= 
            \begin{cases}
                -\infty & \text{if $x - \tfrac{|x|}{\mathit{ed}} - \mathit{em} < \min \mathbb{F}$}, \\
                x - \tfrac{|x|}{\mathit{ed}} - \mathit{em} & \text{otherwise},
            \end{cases} \\
        \RndU x &:= 
            \begin{cases}
                +\infty & \text{if $x + \tfrac{|x|}{\mathit{ed}} + \mathit{em} > \max \mathbb{F}$}, \\
                x + \tfrac{|x|}{\mathit{ed}} + \mathit{em} & \text{otherwise}.
            \end{cases}
    \end{align*}
\end{definition}

For example, $\RndD 0.1 = 0.0855469$ and $\RndU 0.1 = 0.114453$, assuming $\mathbb{F}^+_{4,4}$.
\begin{lemma}
    For $\tilde{x} \in \mathbb{R}$, 
    its rounded value $x \in \mathbb{F}^+_{\mathit{eb},\mathit{sb}}$ with any mode, and
    $\x := [\RndD\tilde{x}, \RndU\tilde{x}]$,
    $\tilde{x} \in \x$ and $v(x) \in \x$ hold.
\end{lemma}


Given an FPA operator $\mathit{op}_\mathbb{F}$ with $n$ arguments,
its \emph{interval extension} $\mathbb{I}^{*n} \to \mathbb{I}^*$ evaluates to intervals enclosing the possible rounded results.
In ordinary IA, interval extensions of real functions are considered (e.g. in \cite{Moore1966,Tucker2011}).
In the same way, we consider interval extension for FPA expressions, but in our case,
handling of ``$\NaN$ cases,'' e.g. $+\infty \times_\mathbb{F} 0$, needs attention.
%
%
In this regard, we will enclose any FPA expressions that may evaluate to $\NaN$ in an interval containing $\NaN$.
Based on the widening operators and the handling of $\NaN$, we define the interval extensions of FPA operators.
\begin{definition}
    Let $\mathit{op}_\mathbb{F}$ be an FPA operator 
    $\mathbb{M}\times\mathbb{F}_{\mathit{eb},\mathit{sb}}^{*n} \to \mathbb{F}_{\mathit{eb},\mathit{sb}}^*$,
    $\mathit{op}_\mathbb{R}$ be an operator $\mathbb{R}^{*n} \to \mathbb{R}^*$ (ideal counterpart of $\mathit{op}_\mathbb{F}$ in RA),
    $\vec{\x}$ be an interval vector in $\mathbb{I}^{*n}$,
    $S$ be the set $\{\mathit{op}_\mathbb{R}(\vec{x}) ~|~ \vec{x}\in\vec{\x}\}$,
    $S_{\setminus\NaN} := S \setminus \{\NaN\}$.
    The \emph{interval extension} of $\mathit{op}_\mathbb{F}$ is defined by 
    \begin{equation*}
        \mathit{op}_\mathbb{I}(\vec{\x}) := [\RndD \inf S_{\setminus\NaN}, \RndU \sup S_{\setminus\NaN}] \cup
        \begin{cases}
            \{\NaN\} & \text{if $\NaN \in S$}, \\
            \emptyset & \text{otherwise}.
        \end{cases}
    \end{equation*}

    Given an FPA expression $f$ that conforms to the syntax category $T$ in Def.~\ref{d:fpa},
    its interval extension is obtained by inductively applying the interval extension to every operator in $f$.
\end{definition}

For example (assuming $\mathbb{F}^+_{4,4}$), 
$[1]\times_\mathbb{I}[0.5] +_\mathbb{I} [0,+\infty] = [0.435547, +\infty]$;
$[0]\times_\mathbb{I}[-\infty,+\infty] = [0] \cup \{\NaN\}$;
$[1]\div_\mathbb{I}[0] = [-\infty, +\infty]$.
Efficient methods to compute $[\RndD \inf S, \RndU \sup S]$ for basic operators, handling only the bounds of the arguments, have been developed for numerical IA libraries; see \cite{Moore1966,Tucker2011}.
In practice, we can have more accurate interval extensions in various ways as long as the resulting intervals are sound,
e.g., we can evaluate $[1]\times_\mathbb{I}[0.5]$ as $[0.5]$.
The following lemma summarizes the soundness of interval extensions.

\begin{lemma} \label{lm:sound}
    Consider an FPA operator $\mathit{op}_\mathbb{F}$ and its interval extension $\mathit{op}_\mathbb{I}$.
    Let $\vec{\x}$ be an interval vector and
    $\f$ be $\mathit{op}_\mathbb{I}(\vec{\x})$.
    We have:
    \begin{align*}
        & \forall m\!\in\!\mathbb{M},~ \forall \vec{x}\!\in\!\vec{\x}, ~
        v( \mathit{op}_\mathbb{F}(m, \vec{x}) ) \in \f.
        %
    \end{align*}
\end{lemma}
The lemma is proved using the \textsc{Why3} tool for the four operators (Appendix~\ref{s:why3}).

\subsection{Approximation of FPA Formulas by IA Formulas}
\label{s:ia:logic}

This section considers \emph{weak} and \emph{strong} abstractions of FPA formulas, based on the interval extensions.
The basic idea here is borrowed from the $\delta$-decision procedure~\cite{Gao2012} that formalizes a numerical process, given a bound $\delta$ for allowed numerical errors.
We apply the idea to the domain of $\mathbb{F}^*$ and do not specify $\delta$ but let the interval-extended operations determine it.

We introduce IA logic formulas in mode weak ($?=-$) or strong ($?=+$). 
\begin{definition}
    Let $?$ be $-$ or $+$ and it is fixed in a formula.
    The grammar of IA formulas, denoted by $\mathbm{\phi}^-$ or $\mathbm{\phi}^+$, is as follows:
    \begin{align*}
        F ~&::=~ T \Rel_?^{[\neg]} T ~|~ F \land F ~|~ F \lor F\\
        T ~&::=~ c ~|~ \mathit{id} ~|~ \mathit{uop}(T) ~|~ \mathit{bop}(T, T) 
    \end{align*}
    where $\Rel_?^{[\neg]}$ is parameterized in three ways: 
    1) $\Rel^{[\neg]}$ represents $\Rel$ or $\Rel^\neg$;
    2) $\Rel \in \{\equiv,=,\geq,>\}$ and $\Rel^\neg \in \{\not\equiv,\neq,<,\leq\}$; and
    3) $\Rel_?$ is instantiated as $\Rel_-$ or $\Rel_+$.
    $c$ and $\mathit{id}$ represent constants (interval literals) and variables in $\mathbb{I}^*$, and $\mathit{uop}$ and $\mathit{bop}$ represent interval operators.
    %
\end{definition}

Modes $-$ and $+$ are prepared for the soundness of decisions of $\Unsat$ and $\Sat$, respectively (Lem.~\ref{lm:ineq} and Th.~\ref{th:dec}).
For the soundness, there is no logical negation operator as in \cite{Gao2012} but we have negated comparison operators in $\Rel_?^\neg$.
Two kinds of operators $\Rel_?$ and $\Rel_?^\neg$ handle ``positive'' and ``negative'' literals separately in the encoding process (Def.~\ref{d:enc}).

The semantics of IA formulas are straightforward, with assignments of free variables in $\mathbb{I}^*$ and evaluating interval extensions.
%
However, in the following, we will modify $\mathbb{I}^*$ slightly to make a sound satisfiability checking.
The interpretation of inequalities $\f \Rel_?^{[\neg]} \g$ in two modes $? \in \{-,+\}$
differs in whether or not to allow uncertain cases such that interval evaluations $\f$ and $\g$ result in non-point intervals and intersect. 
The two groups of operators $\Rel_?$ and $\Rel_?^\neg$ are not only negated but also different in the way they handle $\NaN$.
The semantics of the comparison operators should be appropriately defined so that the following lemma holds.
\begin{lemma} \label{lm:ineq}
    Consider the following subset of $\mathbb{I}^*$:
    \begin{equation*}
        \mathbb{I}^*_\# := \{\x\!\in\!\mathbb{I}^{*} ~|~ \exists \hat{x}\!\in\!\mathbb{F}^*_{\mathit{eb},\mathit{sb}},~ \hat{x}\in\x\}.
    \end{equation*}
    Let $\f$ and $\g$ be interval extensions of $m$-ary and $n$-ary FPA expressions $f$ and $g$, respectively. We have:
    \begin{align*}
        \text{$\f \Rel_- \g$ is $\Unsat$} &~\Rightarrow~ \text{$f \Rel_\mathbb{F} g$ is $\Unsat$}, \\
        \text{$\f \Rel_-^\neg \g$ is $\Unsat$} &~\Rightarrow~ \text{$\neg(f \Rel_\mathbb{F} g)$ is $\Unsat$}, \\
        \exists (\vec{\x},\vec{\y}) \in \mathbb{I}^{* m+n}_\#,~ \f(\vec{\x}) \Rel_+ \g(\vec{\y}) 
        &~\Rightarrow~ \text{$f \Rel_\mathbb{F} g$ is $\Sat$}, \\
        \exists (\vec{\x},\vec{\y}) \in \mathbb{I}^{* m+n}_\#,~ \f(\vec{\x}) \Rel_+^\neg \g(\vec{\y}) 
        &~\Rightarrow~ \text{$\neg(f \Rel_\mathbb{F} g)$ is $\Sat$}.
    \end{align*}
\end{lemma}
The lemma is proved using \textsc{Why3} to confirm that every comparison operators are correctly defined, but for limited forms of $\f$ and $\g$ (Appendix~\ref{s:why3}).

Because assignments with intervals that do not contain any FP numbers (e.g. $[0.1]$) are possible, we must prohibit them in $\mathbb{I}^*_\#$ to make a sound decision for strong interval extension.
In an actual encoding, the condition ``$\exists \hat{x}\!\in\!\mathbb{F}^*_{\mathit{eb},\mathit{sb}}, {\hat{x}\in\x}$'' can be made simpler and weaker, e.g., as $\LB{x} \leq \RndD\UB{x}$ or $\RndU\LB{x} \leq \UB{x}$.
In the decision with weak interval extension,
it is sufficient to assume only point intervals (and point intervals appended with $\{\NaN\}$)
because any FP number can be represented by a point interval;
in addition, an evaluation with point intervals will give the best approximation.


As an example of the operators, instances of $>_?^{[\neg]}$, which are $>_-$, $\leq_-$ (i.e. $>_-^\neg$), $>_+$ and $\leq_+$ (i.e. $>_+^\neg$), when rhs is $[0]$ are defined in a logic on $\mathbb{R}^*$ as follows:
\begin{align*}
    \f >_- [0]          &~:\Leftrightarrow~ \UB{f}_{\setminus \NaN} > 0, & 
    \f >_+ [0]          &~:\Leftrightarrow~ \NaN \not\in \f \land \LB{f} > 0, \\
    \f \leq_- [0]       &~:\Leftrightarrow~ \NaN\in\f \lor \LB{f}_{\setminus\NaN} \leq 0, &
    \f \leq_+ [0]       &~:\Leftrightarrow~ \UB{f}_{\setminus\NaN} \leq 0,
\end{align*}
where $\f_{\setminus\NaN}$ denotes $\f \setminus \{\NaN\}$.
Since $\NaN >_\mathbb{F} 0$ does not hold, 
$\NaN$ cases are disallowed by $>_+$ for the soundness, whereas 
they are ignored by $>_-$ for the completeness for the case where $f$ is not $\NaN$.
On the other hand, since the negative literal $\neg(\NaN >_\mathbb{F} 0)$ holds, $\NaN$ cases are handled differently by $\leq_?$.
The full definition of the operator semantics is described in Appendix~\ref{s:compop}.


Let $\f$ be an interval extension of $f$, $\delta$ be the width of $\f$ (i.e. $\UB{f}-\LB{f}$), and $\delta_\mathbb{F}$ be the upward rounded value of $\delta$ in $\mathbb{F}_{\mathit{eb},\mathit{sb}}$.
When contrasted with the $\delta$-decision procedure~\cite{Gao2012}, 
checking the satisfiability of $\f \Rel_- [0]$ is equivalent to checking whether 
$f \Rel -\delta_\mathbb{F}$ 
is satisfiable or $f \Rel 0_\mathbb{F}$ is not satisfiable;
likewise, checking $\f \Rel_+ [0]$ (with the above conditioning) is equivalent to checking whether $f \Rel 0_\mathbb{F}$ is $\Sat$ or 
$f \Rel \delta_\mathbb{F}$ 
is $\Unsat$.
%

Next, we consider translation from FPA into IA.
To encode FPA, 
some expressions in RA are also used to describe boundary conditions of intervals.

\begin{definition} \label{d:enc}
    The \emph{weak extension} $\mathbm{\phi}^-$ or \emph{strong extension} $\mathbm{\phi}^+$ is translated from an FPA formula $\phi$ by the following steps:
    \begin{enumerate}
        \item Transform $\phi$ into a negation normal form.
        \item Transform each literal into an interval inequality of the form $\f \Rel_?^{[\neg]} \g$;
            positive (resp. negative) literals are encoded using operators $\Rel_?$ (resp. $\Rel_?^\neg$), e.g., $f < g$ into $\g >_? \f$ and $\neg (f < g)$ into $\g \leq_? \f$.
            Other than that, translation is straightforward (constants to point intervals, operators to their interval extensions, etc.).
        \item When $? = +$, each variable $x$ in $\phi$ is translated into an interval variable $\x$, appended with a constraint $\LB{x} \leq \RndD \UB{x}$ (or $\RndU\LB{x} \leq \UB{x}$).
            When $? = -$, variables are forced to be a point interval with constraint $\LB{x} = \UB{x}$.
    \end{enumerate}
\end{definition}
    
From the above definitions and lemmas, the following theorem holds.
\begin{theorem} \label{th:dec}
    Let $\phi$ be an FPA formula and $\mathbm{\phi}^-$ and $\mathbm{\phi}^+$ be weak and strong interval extensions of $\phi$, respectively.
    \begin{itemize}
    \item If $\mathbm{\phi}^-$ is not satisfiable, then so is $\phi$.
    \item If $\mathbm{\phi}^+$ is satisfiable, then so is $\phi$.
    \end{itemize}
\end{theorem}


For example, consider an $\Unsat$ FPA formula $\phi :\Leftrightarrow x >_\mathbb{F} 0 \land -x >_\mathbb{F} 0$; 
$\mathbm{\phi}^-$ is not satisfiable because no point intervals satisfy the two predicates;
$\mathbm{\phi}^+$ is also not satisfiable because no interval $\f$ satisfies both $\LB{f} > 0$ and $-\UB{f} > 0$ (cf. the definition of $\f >_+ [0]$).
An FPA formula $\phi' :\Leftrightarrow \neg(x >_\mathbb{F} 0) \land \neg(-x >_\mathbb{F} 0)$ is satisfiable with the assignment $x := 0_\mathbb{F}$ or $x := \NaN$.
Its interval extensions $\mathbm{\phi}'^?$ are of the form $\x \leq_? [0] \land -\x \leq_? [0]$ (constraint is also appended to $\mathbm{\phi}'^+$ in Step~3);
then, $\mathbm{\phi}'^-$ is satisfiable with $\x := [0]$ or any $\x$ containing $\NaN$;
$\mathbm{\phi}'^+$ is not satisfiable because the auxiliary constraint forbids $\x := [0]$.

%

\section{Implementation}
\label{s:impl}

\begin{figure}
    \centering
    \includegraphics[width=.8\textwidth]{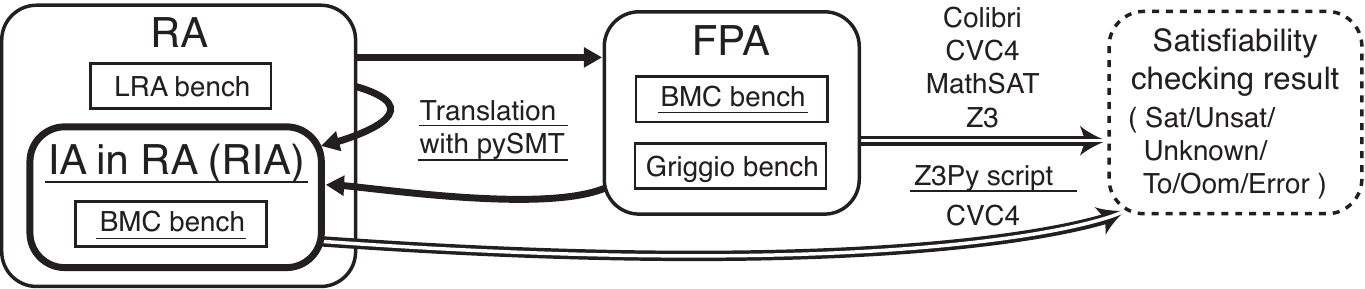} 
    \caption{\label{f:proc} Process of experiments. Underlined items are newly developped.}
\end{figure}

We have implemented a solver for FPA formulas via translation into weak and strong interval extensions;
our implementation expresses IA formulas in real arithmetic (RA) and solves them using an SMT solver (we use \textsc{CVC4} and \textsc{Z3}).
In addition, we have prepared several benchmark problems for the experiments (Sect.~\ref{s:xp}).
The process is illustrated in Fig.~\ref{f:proc}.
In the following, we denote ``IA embedded in RA'' by RIA.
The main process of the proposed solver is twofold: 1) Translation from FPA to RIA; 
2) An incremental solving process in which the FPA precision is gradually improved to accelerate the overall process.
For benchmarking, we prepared two sets of problems in FPA; also, we prepared a set by translating problems in RA into FPA or RIA.

\subsection{Encoding IA in RA}
\label{s:impl:enc}

\begin{figure}[p]
    \lstset{frame=single}
    \lstset{numbers=left}
    \begin{lstlisting}[basicstyle=\ttfamily\footnotesize]
;; %Definition of datatype RInt representing $\mathbb{I}^*$.%
(declare-datatype RInt ((tpl (ri.l Real) (ri.u Real) (p_nan Bool) )))

;; ...
;; %Definition of $\RndD(v)$.%
(define-fun ri.r_dn ((v Real)) Real
  (let ((w (- v (/ (ite (>= v 0) v (- v)) %$\mathit{ed}$%) %$\mathit{em}$%)))
    (ite (>= w (- ri.max_value)) w (- ri.large_value)) ) )

;; ...
;; %Definition of $\x + \y$.%
(define-fun ri.add ((x RInt) (y RInt)) RInt
  (let ( (l (ri.r_dn (+ (ri.l x) (ri.l y)))) 
         (u (ri.r_up (+ (ri.u x) (ri.u y)))) )
    (tpl l u (or (p_nan x) (p_nan y) (and (is_ninf x) (is_pinf y)) 
                     (and (is_pinf x) (is_ninf y)) )) ) )

;; ...
;; %Definition of $\x \times \y$.%
(define-fun ri.mul ((x RInt) (y RInt)) RInt
  (ite (>= (ri.l x) 0)
    (ite (= (ri.u x) 0)
      (ite (and (not (is_ninf y)) (not (is_pinf y)) 
                    (not (p_nan x)) (not (p_nan y)) )
        ri.zero ;; [x] = [0]
        ri.zero_nan ) ;; [x] = [0] and [y] = -+inf
      (ite (>= (ri.l y) 0)
        (ite (= (ri.u y) 0)
          ;; Other 18 cases are omitted.
        ) ) ) ) )

;; ...
;; %Definitions of $\f >_- [0]$ and $\f >_- \g$.%
(define-fun ri.gt0 ((f RInt)) Bool
  (or (is_pinf f) (> (ri.u f) 0)) )

(define-fun ri.gt ((f RInt) (g RInt)) Bool
  (or (is_pinf f) (is_ninf g) (ri.gt0 (ri.sub_exact f g))) )

;; ...

(declare-const x RInt)
(assert (= (ri.l x) (ri.u x)))
(assert (=> (p_nan x) (= x ri.nan)))

(assert (ri.gt (ri.mul (ri.of_real (/ 1 10)) x) (ri.exact 1.0)))
    \end{lstlisting}
    \caption{\label{c:ex} Example of IA encoding in RA ($?=-$).}
\end{figure}

Given an FPA formula $\phi$, our translator generates RIA descriptions that encode $\mathbm{\phi}^-$ and $\mathbm{\phi}^+$.
As long as $\phi$ consists of linear expressions, the translation is done in linear RA.
Example translation from an FPA formula $\times(\mathrm{RNE}, 0.1_\mathrm{RNE}, x) >_\mathbb{F} 1$ is shown in Fig.~\ref{c:ex}, where $0.1_\mathrm{RNE}$ is a rounded value with mode RNE.

In the beginning, \textbf{Lines~1--38} defines the vocabularies of RIA.
At \textbf{Line~2}, we prepare the \lstinline{RInt} datatype to represent intervals, defined as tuples of the bounds and a flag indicating whether $\NaN$ is contained.
At \textbf{Lines~12--16}, the downward rounding operator is defined following Def.~\ref{d:rounding}.
Placeholders $\mathit{ed}$ and $\mathit{em}$ should be filled with concrete values.
The symbol \lstinline{ri.max_value} represents the maximum representable number prepared for the considered FP sort, and 
\lstinline{ri.large_value} is constrained as $\texttt{ri.large\char`_value} > 2\ \texttt{ri.max\char`_value}$ and is used to represent $\infty$.
At \textbf{Lines 12--30}, interval operators are defined following a typical algorithm, e.g. \cite{Ishii2020}, making case analyses on the bounds of argument intervals.
If a $\NaN$ case may be involved, the functions compute the bounds of a normal interval obtained for the other cases and set the flag \lstinline{p_nan};
for example, the branch at \textbf{Line~26} might involve $\NaN$ cases, i.e., $x$ or $y$ is $\NaN$, or $0 \times \mp\infty$, so it results in the interval $[0]\cup\{\NaN\}$.
At \textbf{Lines~34--38}, definitions of comparison operators follow the discussion in Sect.~\ref{s:ia:logic}.
In the definition of function \lstinline{ri.gt},
operator \lstinline{ri.sub_exact} is used for subtraction without widening the resulting interval.

Finally, at \textbf{Lines~42--46}, the example formula is specified.
The variable $x$ is declared with auxiliary constraints,
i.e., $\LB{x}=\UB{x}$ (cf. Step~3 of Def.~8) and a constraint for the $\NaN$ assignment.%
\footnote{For simplicity of encoded formulas, we have chosen not to handle the interval $\{\NaN\}$; instead, we assign the value $[-\infty]\cup\{\NaN\}$ (for $\mathbm{\phi}^-$) or $[-\infty,+\infty]\cup\{\NaN\}$ (for $\mathbm{\phi}^+$).}

\emph{Multi-precision encoding scheme.}
To encode formulas involving multi-precision FP numbers, we use a modified encoding scheme.
It assumes a list of precisions $(\mathit{eb}_i,\mathit{sb}_i)$ that appear in a formula (each precision is represented by an integer $i$).
Then, the scheme uses a set of rounding operators prepared for each precision and modified operator functions with an additional precision parameter.

\subsection{Translators}
\label{s:impl:trans}

We have implemented a translator from FPA descriptions to RIA descriptions.
It is realized by extending the implementation of \textsc{pySMT},\footnote{\url{https://github.com/pysmt/pysmt}.}
a \textsc{Python} library for the SMT format containing a parser, printers, etc.
We implemented support for FPA, intermediate representation of vocabularies of IA, and translation and printing scripts. Embedding in RA was implemented in the printers.
The translator runs in several ways e.g. for weak or strong mode.
It can also generate formulas in which precisions are abstracted for incremental solving.
To facilitate the experiments in Sect.~\ref{s:xp},
we have also implemented translators from RA to FPA and RIA.
The implementation is available at \url{https://github.com/dsksh/pysmt}.

\subsection{Solver Script}
\label{s:impl:solver}

We have implemented a \textsc{Python} script to solve RIA formulas.
The script runs two processes for mode $-$ or $+$ in parallel; it results in $\Unsat$ or $\Sat$ if either of the processes obtains a sound result; otherwise, it results in $\Unknown$.
The script is based on \textsc{Z3Py}~2.8.12\footnote{\url{https://github.com/Z3Prover/z3}.}.
In addition to default solving process, the script implements incremental process, which tries to solve under several precisions of FP numbers from coarser to exact ones (Appendix~\ref{s:impl:inc}).

\section{Experiments}
\label{s:xp}

We have conducted experiments to answer the following questions:
(\textbf{RQ1})~How efficient is the proposed method when compared to the state-of-the-art FPA solvers? 
(\textbf{RQ2})~To what extent does the incompleteness of Alg.~\ref{a:inc} affect the results in practice?
We have experimented using three sets of problem instances.

In the experiments, we solved FPA formulas via encoding into RIA.
Formulas were then solved in three ways:
1) Using the solver script (Sect.~\ref{s:impl:solver}) with non-incremental setting;
2) With incremental setting;
3) Using \textsc{CVC4}~1.8\footnote{\url{https://cvc4.github.io}.} with manual selection of conclusive results.
We refer to our method with either of the settings 1--3 as ``RIA.''
For comparison, we also solved with the exiting FPA solvers \textsc{Z3}, \textsc{CVC4} (linked with \textsc{SymFPU}~\cite{Brain2019}), \textsc{Colibri}~v2176~\cite{Marre2017}, and \textsc{MathSAT}~5.6.6\footnote{\url{https://mathsat.fbk.eu}.} (with an ACDCL-based FPA solver enabled).
Experiments were run on a 2.2GHz Intel Xeon E5-2650v4 
with a memory limit of 3GB.
The timeout was set to 1200s.
We did not measure the time taken for translation, but only the time taken for the solving process for FPA or RIA formulas.

\subsection{Bounded Model Checking}
\label{s:xp:bmc}

\begin{figure}[t!]
    \centering
    \subfloat[Width of error bounds $\Delta^-+\Delta^+$.]{%
        \begin{minipage}[b]{.2\textwidth}
            \hspace*{1em}
            \includegraphics[width=.8\textwidth]{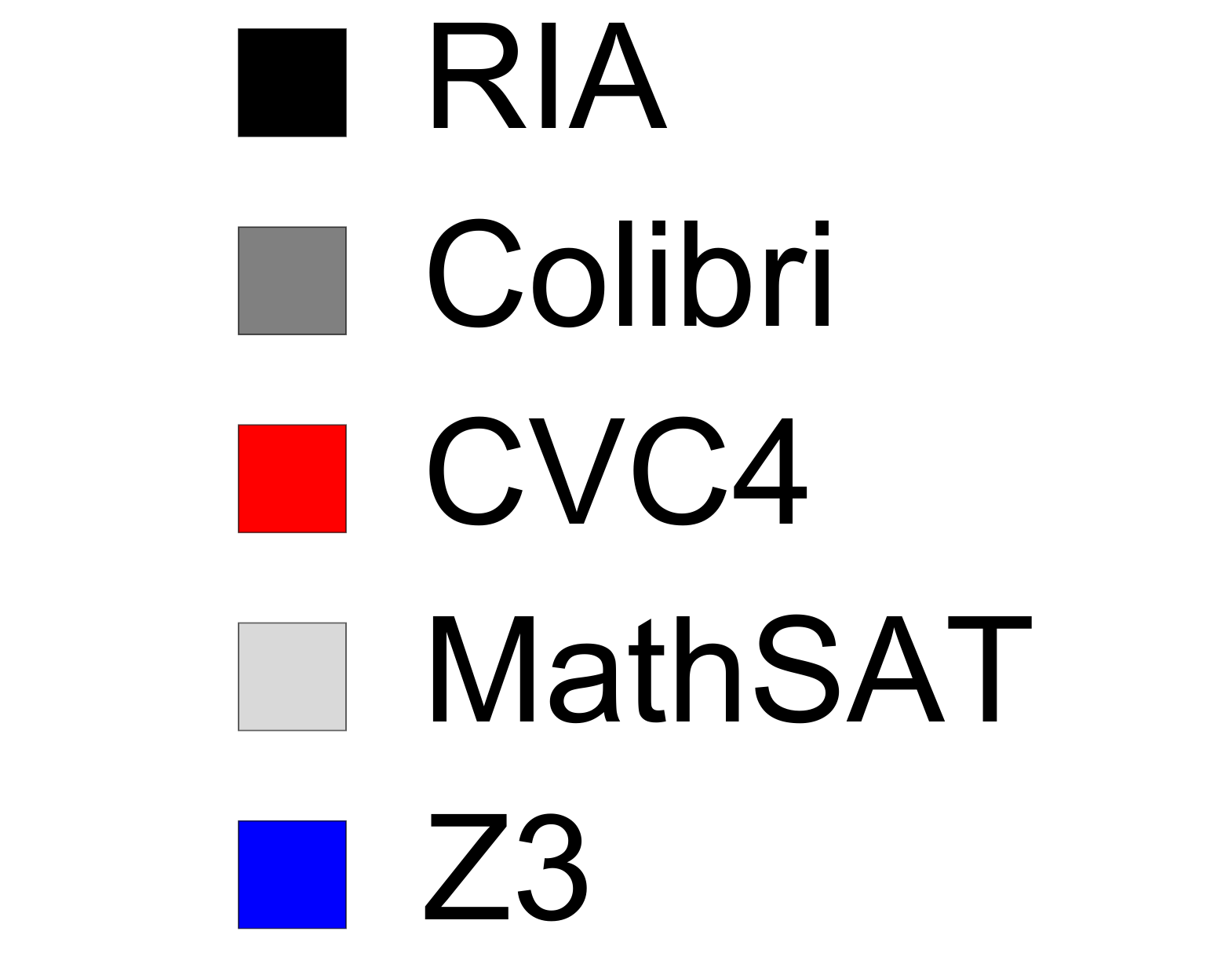}

            \vspace*{1em}

            \hspace*{-1em}
            \includegraphics[width=1.1\textwidth]{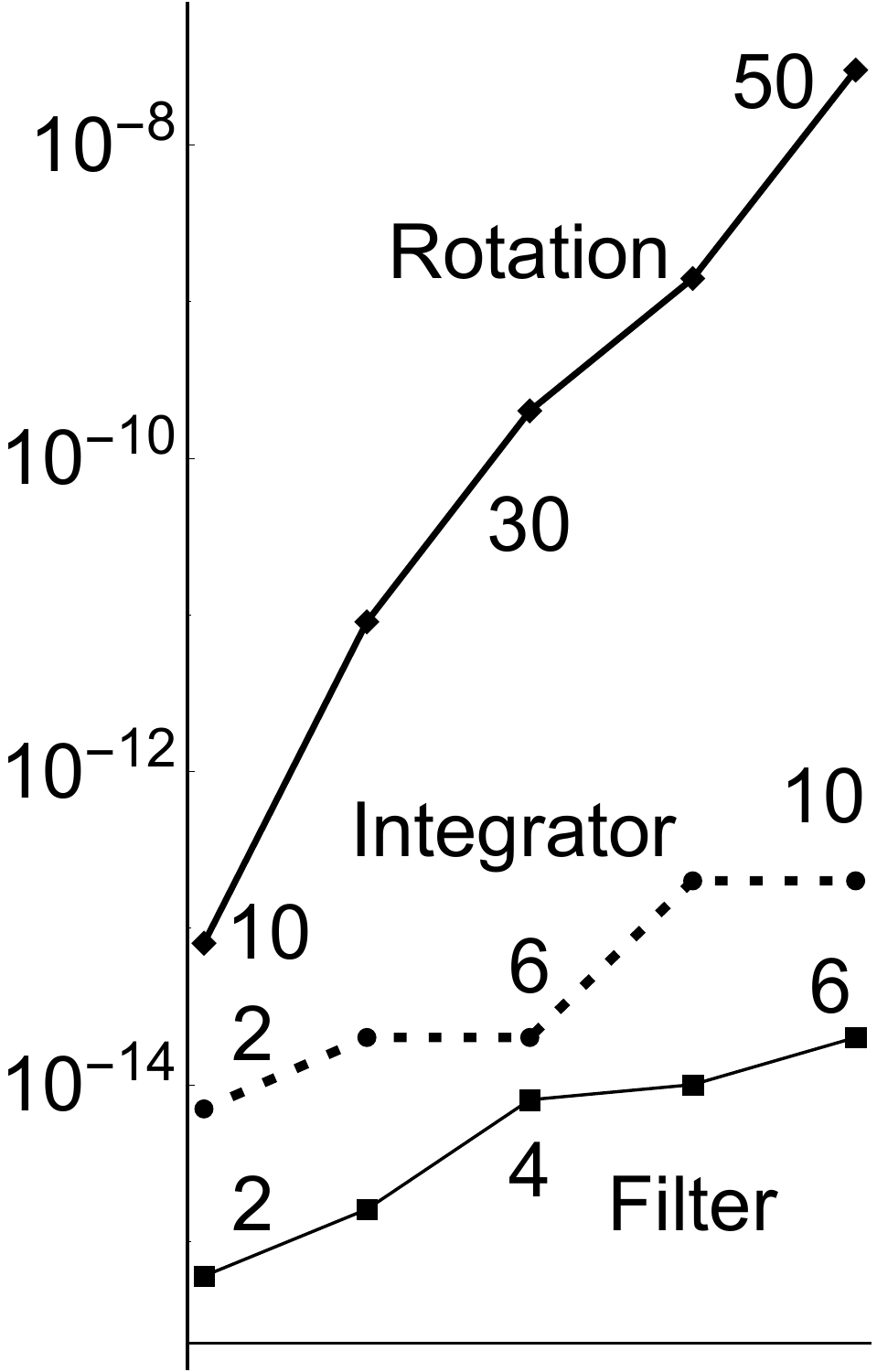}

            \vspace*{.5em}
        \end{minipage}
    }
    \subfloat[Integrator.]{%
        \begin{minipage}[b]{.25\textwidth}
            \includegraphics[height=.17\textheight]{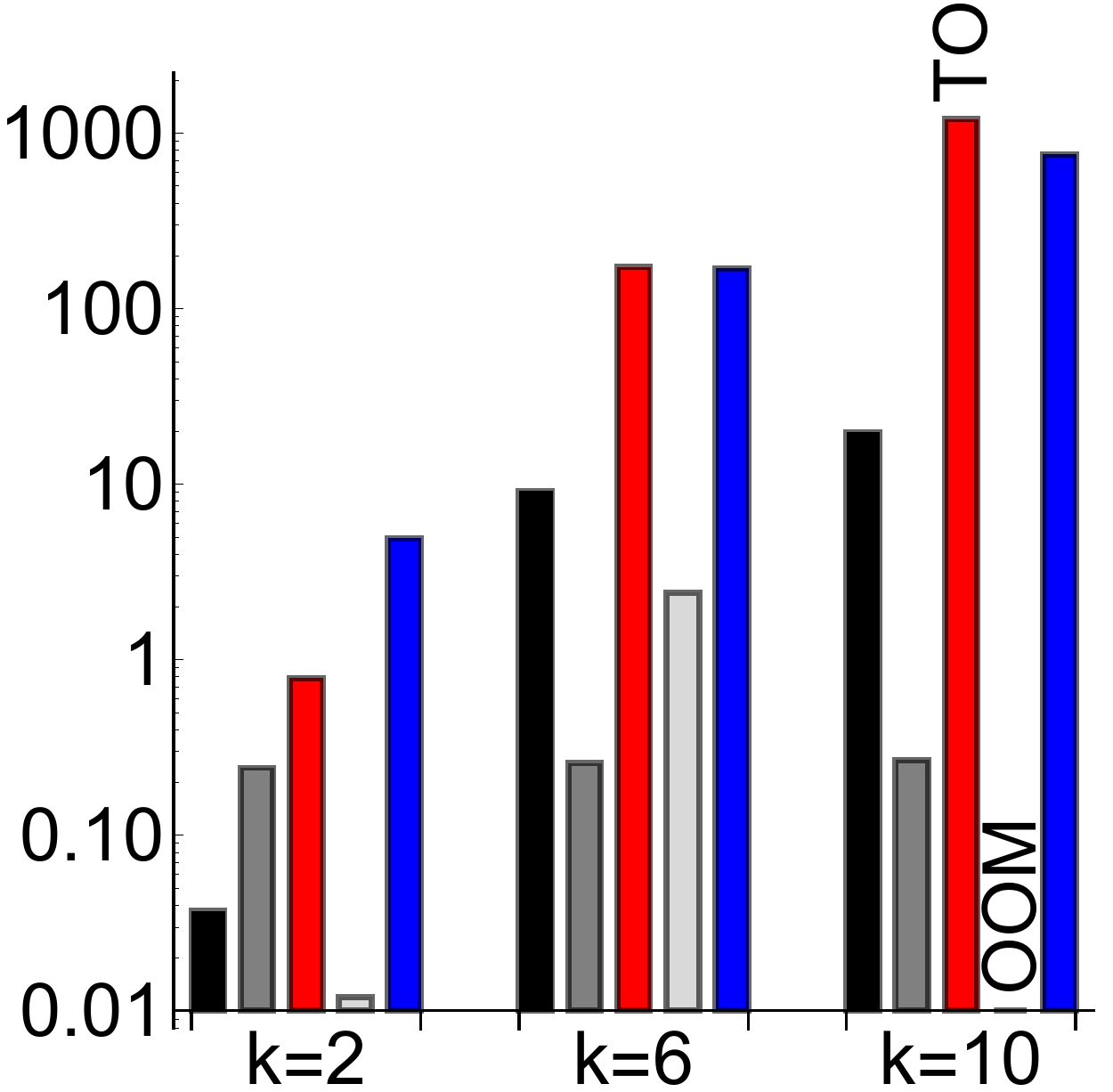} 
            \includegraphics[height=.17\textheight]{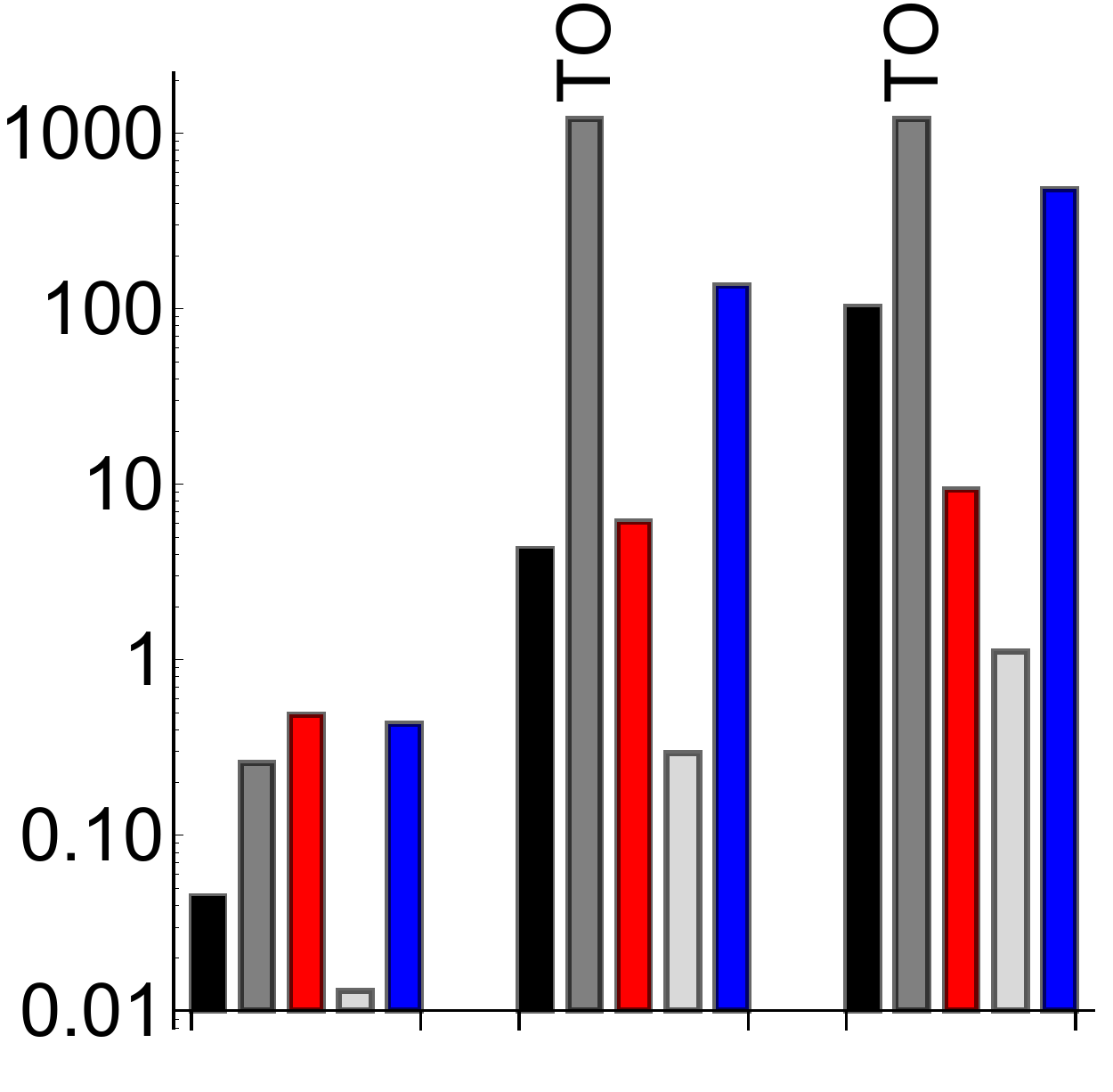}
        \end{minipage}}
    \subfloat[Filter.]{%
        \begin{minipage}[b]{.25\textwidth}
            \includegraphics[height=.17\textheight]{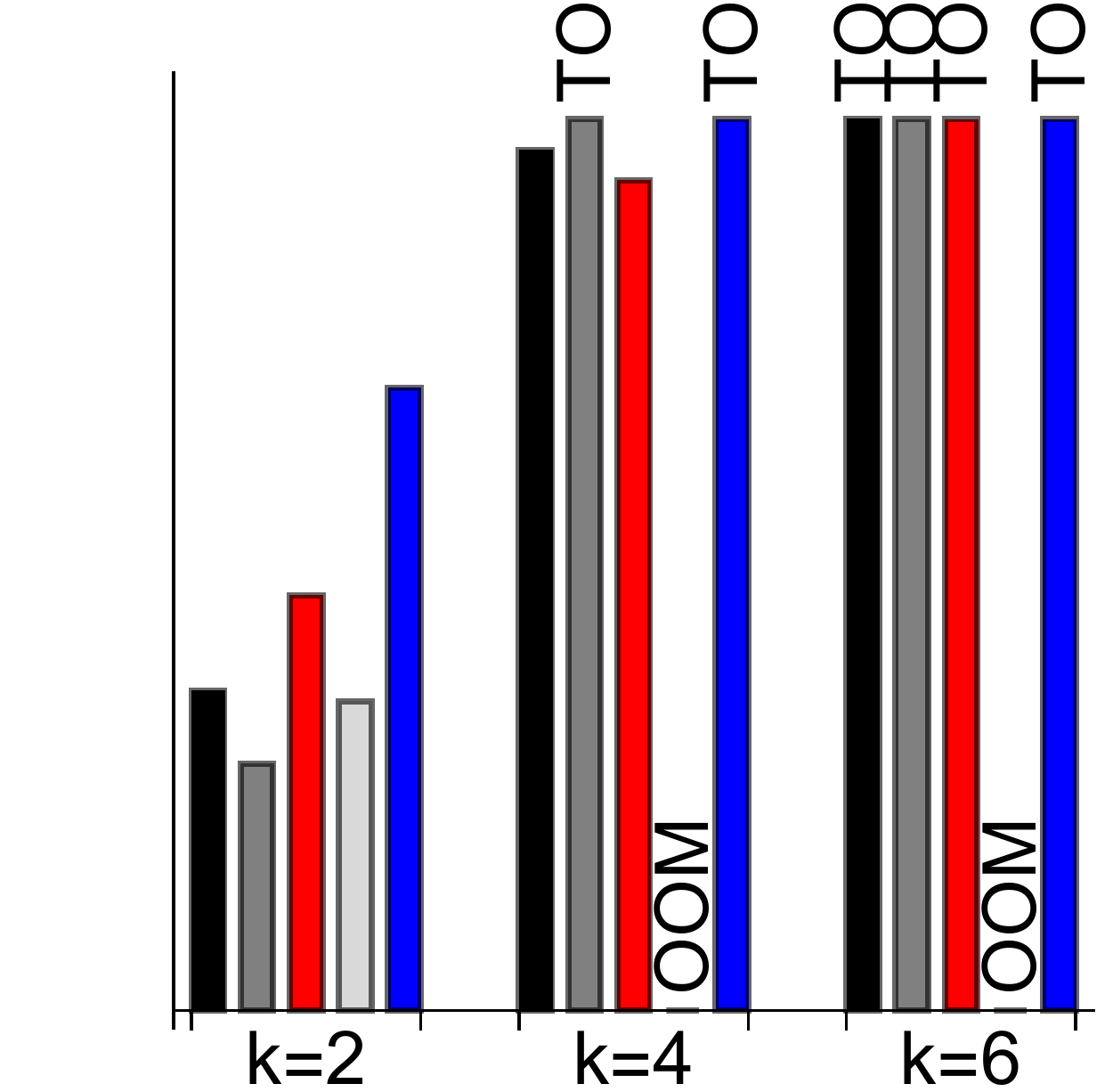} 
            \includegraphics[height=.17\textheight]{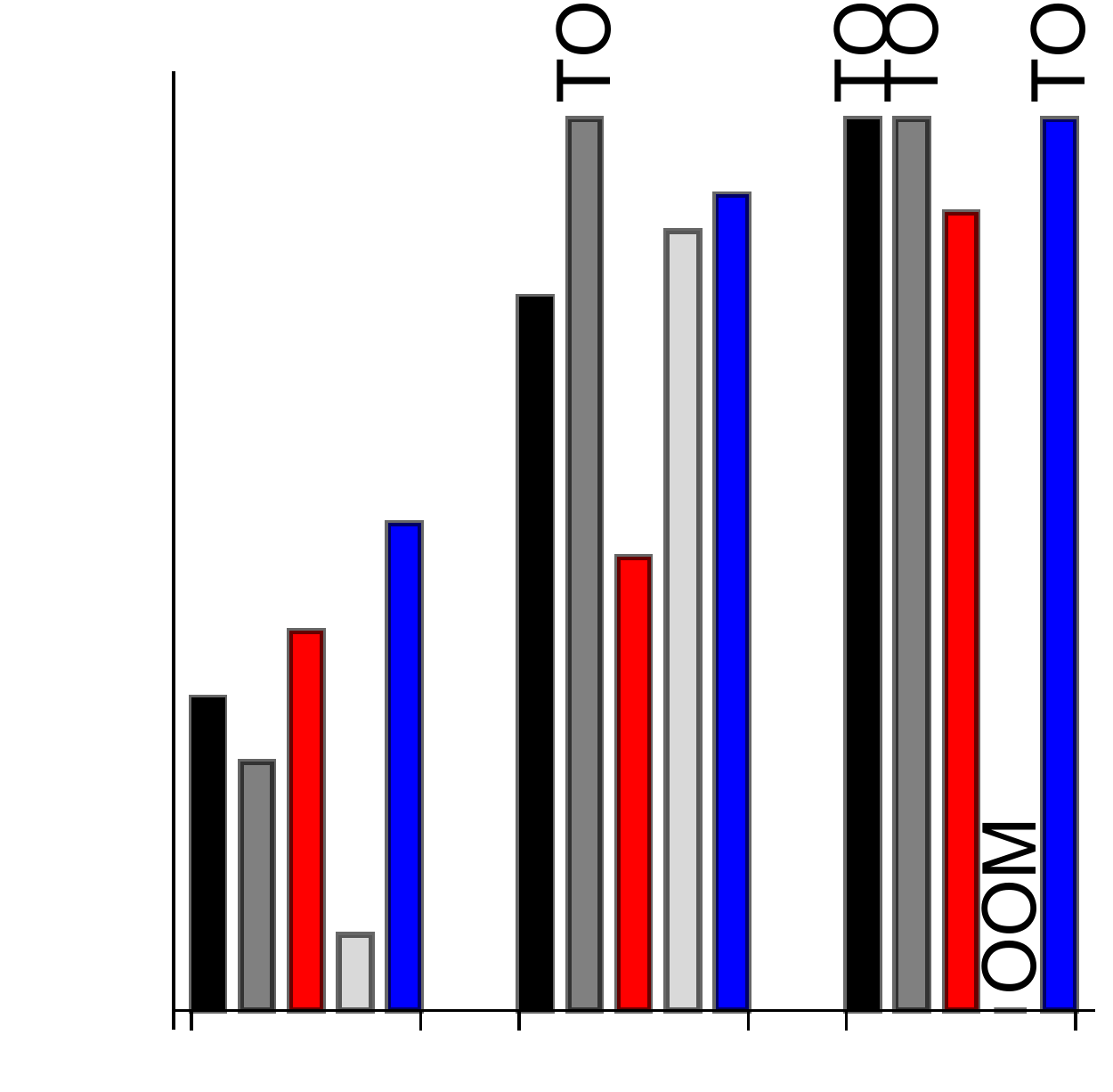}
        \end{minipage}}
    \subfloat[Rotation.]{%
    \begin{minipage}[b]{.25\textwidth}
        \includegraphics[height=.17\textheight]{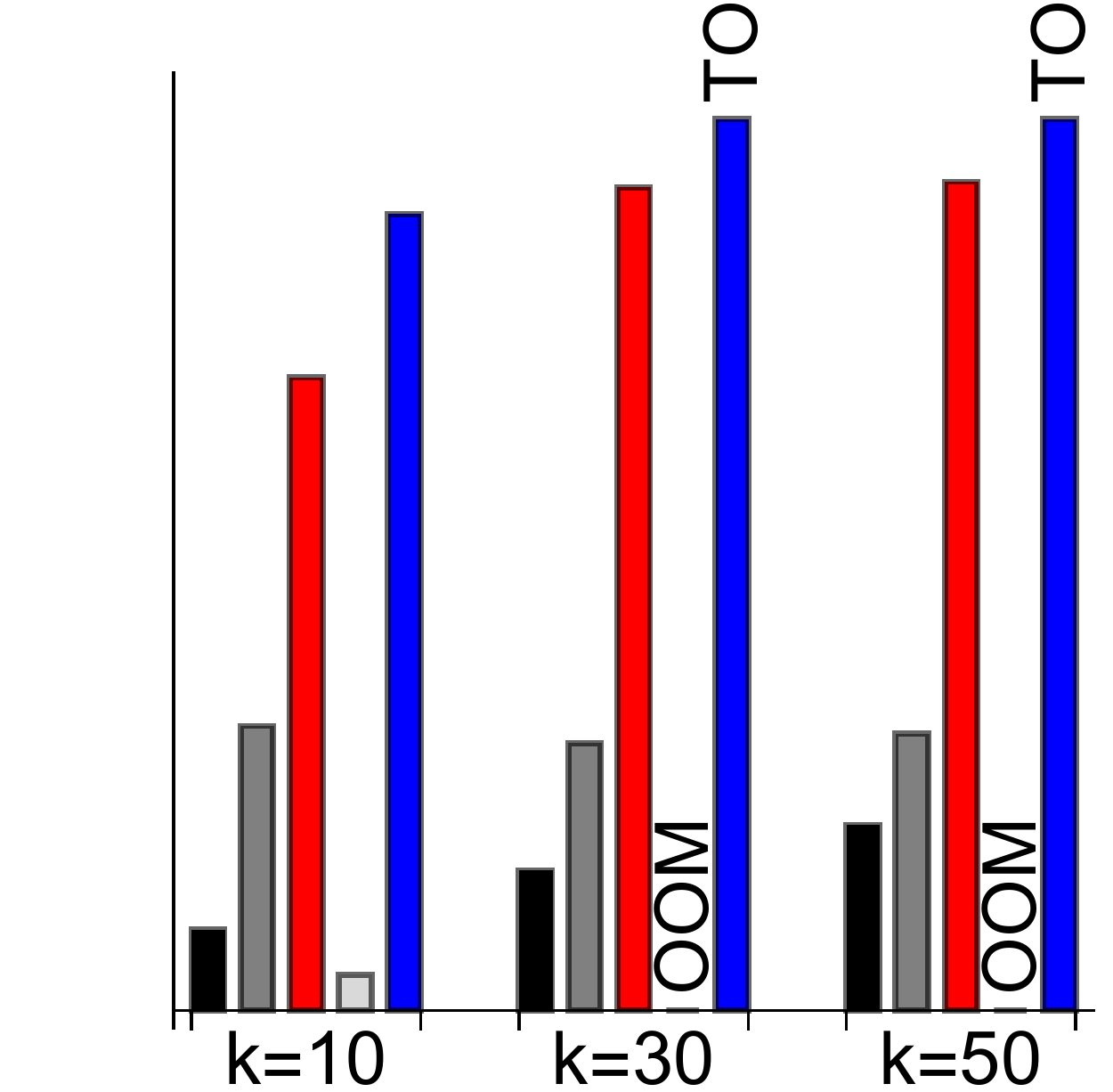} 
        \includegraphics[height=.17\textheight]{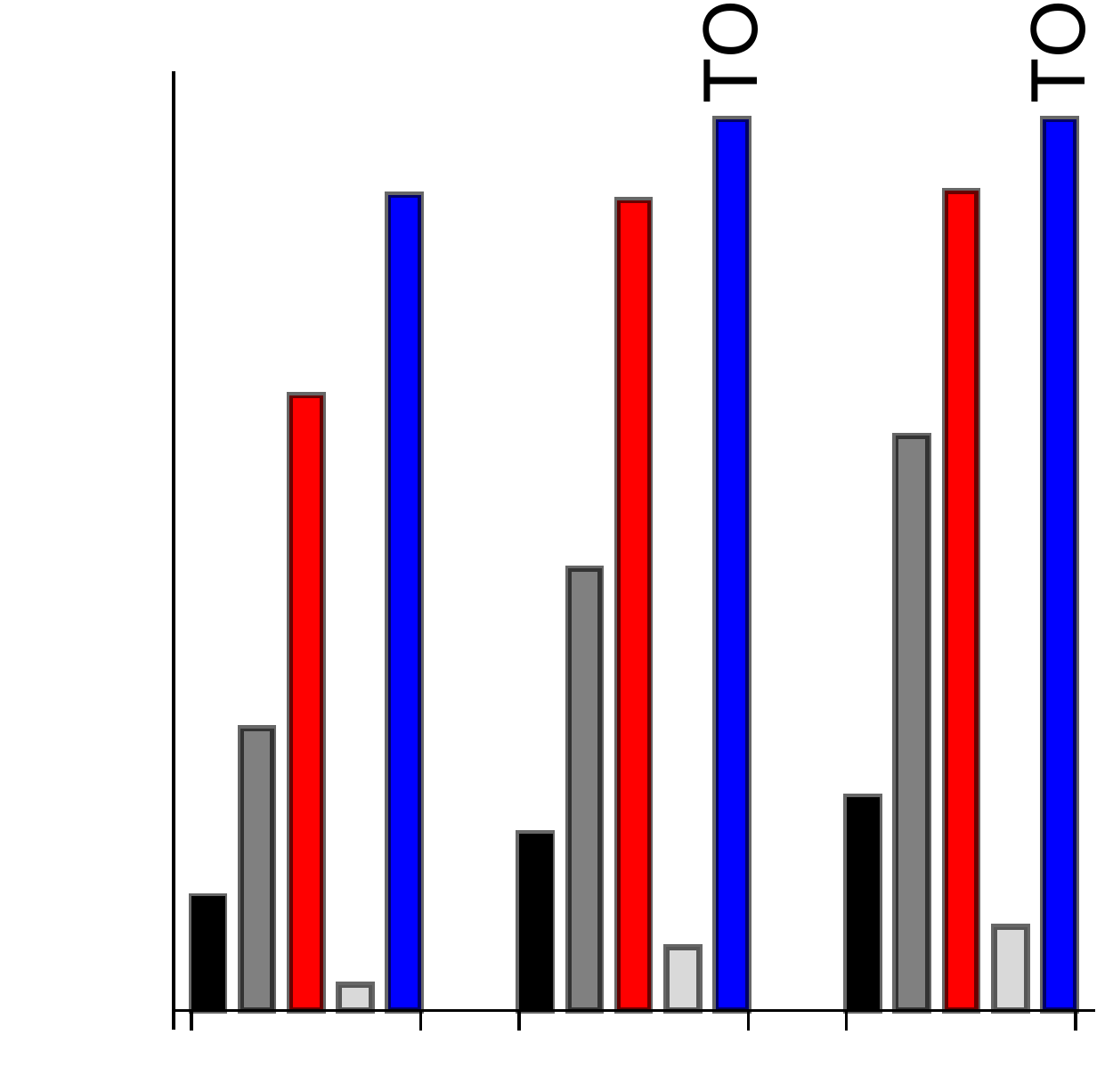}
    \end{minipage}}
    \vspace{-.7em}
    \caption{\label{f:bmc} Results of BMC. 
    Chart~(a) shows required perturbations to make each BMC conclusive.
    (b)--(d) show the execution time for $\Unsat$ (upper) and $\Sat$ (lower) instances. ``TO'' and ``OOM'' represent executions resulted in timeout and out of memory.}
\end{figure} 

In the first experiment,
we performed the bounded model checking (BMC) of discrete-time dynamical systems as a practical use case.
In BMC with a bound $k \in \mathbb{N}$, paths of a target system of length $k$ were encoded into an FPA formula $\phi$ in $\mathbb{F}_{11,53}$ (rounding modes were left unspecified), and we verified whether an output $o$ of a path reaches a threshold $\mathit{th}$ by checking the satisfiability of $\phi \land {o \geq \mathit{th}}$.
As target systems, we considered a 1D feedback integrator, a 2D second-order filter, and a rotation on a 2D plane (see Appendix~\ref{s:xp:bmc:spec} for their specifications); 
a transition of the systems involves 2, 5 or 6 arithmetic operations, respectively.

For each system, we performed BMC with three $k$s.
We checked for each system and $k$ a boundary threshold value $\widetilde{\mathit{th}}$ whose perturbation switches the satisfiability.
We then obtained for each instance the \emph{error bounds} $\Delta^- < 0$ and $\Delta^+ > 0$ such that the proposed method outputs $\Unsat$ or $\Sat$ when the threshold is $\mathit{th} := \widetilde{\mathit{th}}-\Delta^?$.
Finally, we solved the RIA or FPA formulas encoding the $\Unsat$ and $\Sat$ instances perturbated for $\Delta^-$ and $\Delta^+$.
We compared the execution time of the prepared solvers;
wherein, the non-incremental RIA solver (setting 1) was used to have best results since every instance requires fine precision.
The experimental results are shown in Fig.~\ref{f:bmc}.

\subsection{Benchmark Problems}
\label{s:xp:bench}

\begin{figure}[t!]
    \centering
    \subfloat[LA (RIA is with setting 3).]{%
        \begin{minipage}[b]{.55\textwidth}
            \raisebox{5em}{\includegraphics[width=.4\textwidth]{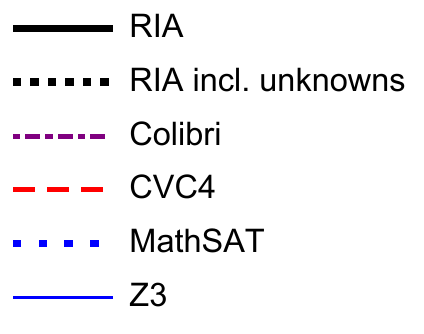}}
            \hspace*{-5em}
            \includegraphics[height=.2\textheight]{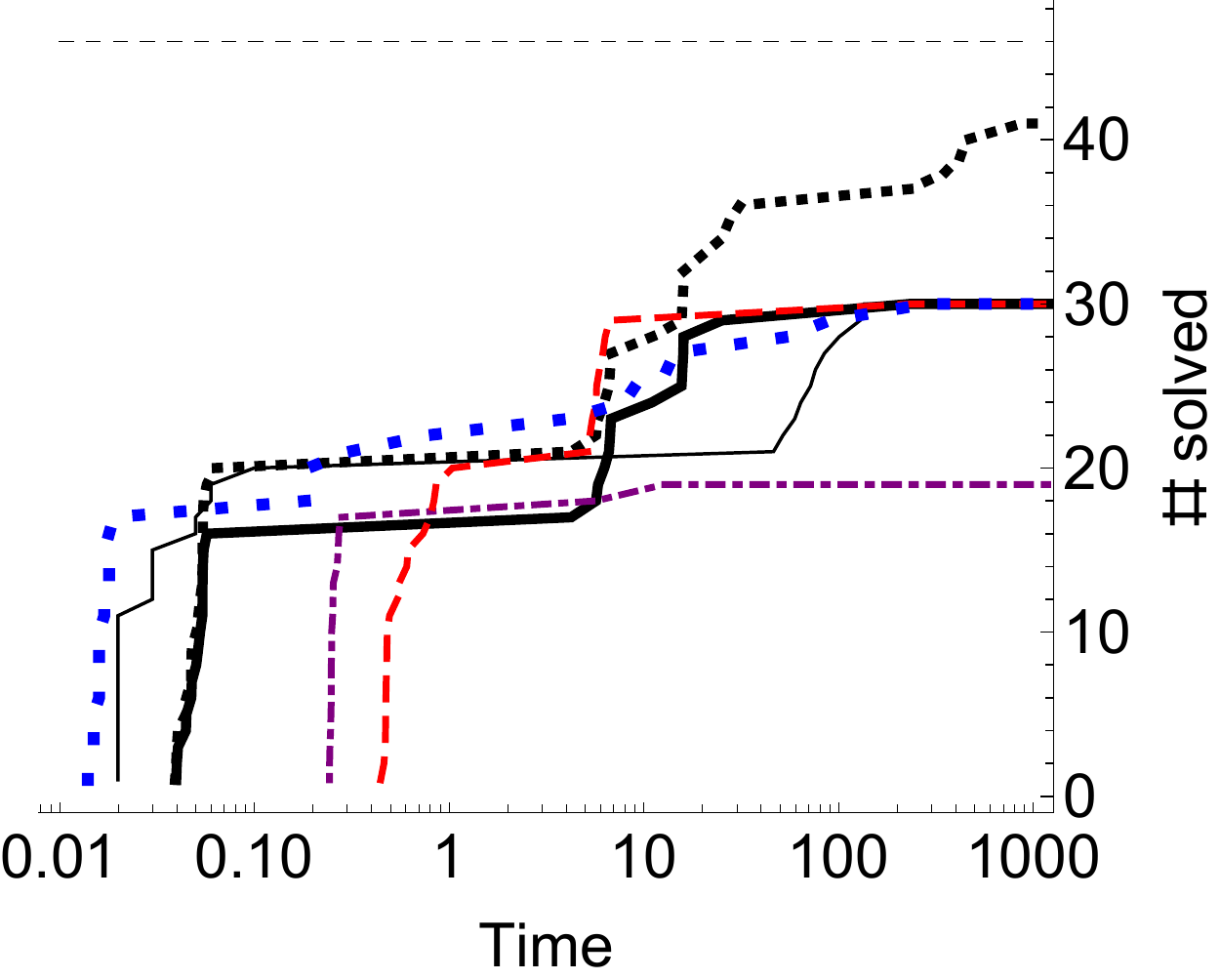} 

            \hspace*{4em}
            \includegraphics[height=.2\textheight]{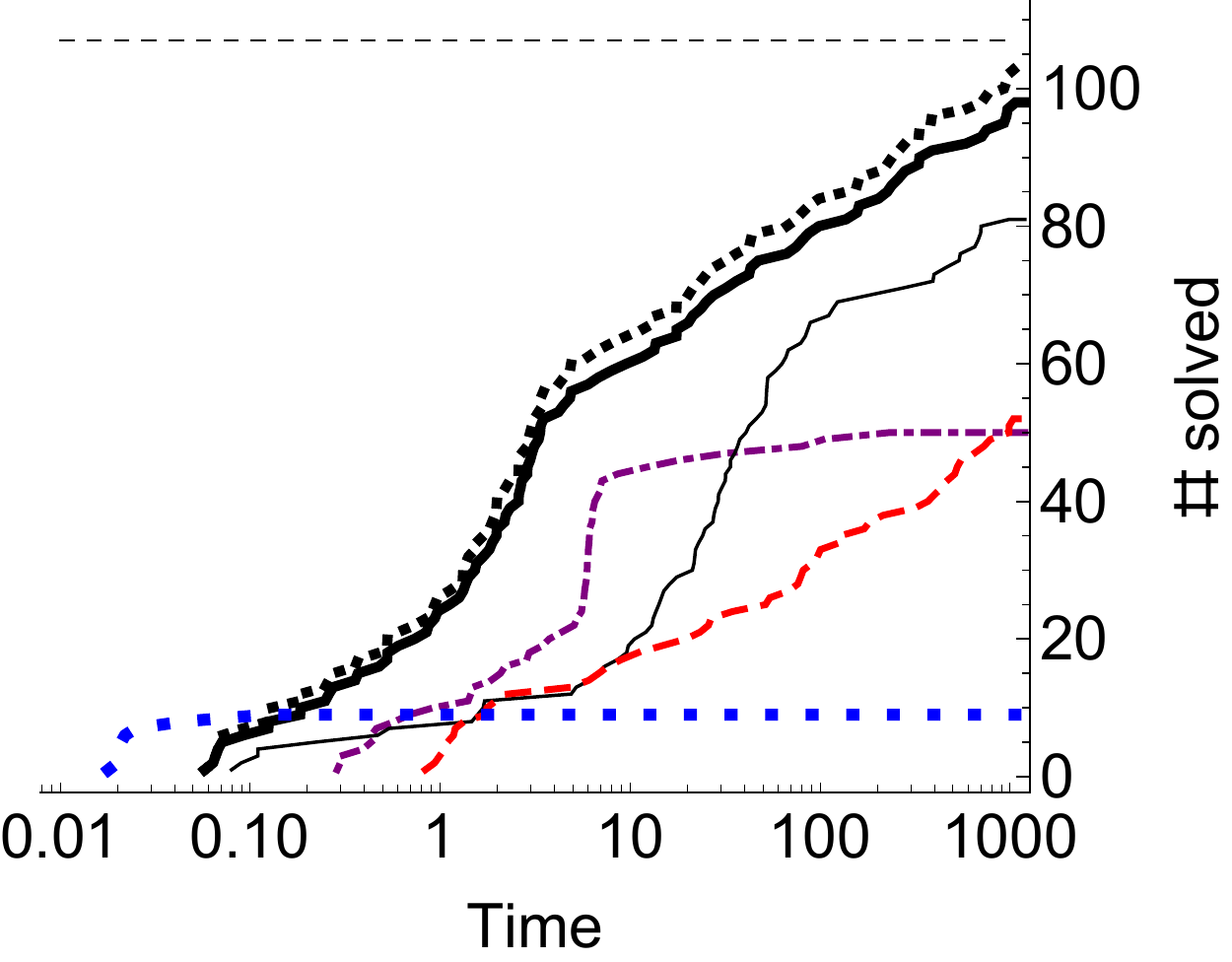} 
        \end{minipage}}
    \subfloat[Griggio (RIA is with setting 2).]{%
        \begin{minipage}[b]{.45\textwidth}
            \hspace*{1.5em}
            \includegraphics[height=.2\textheight]{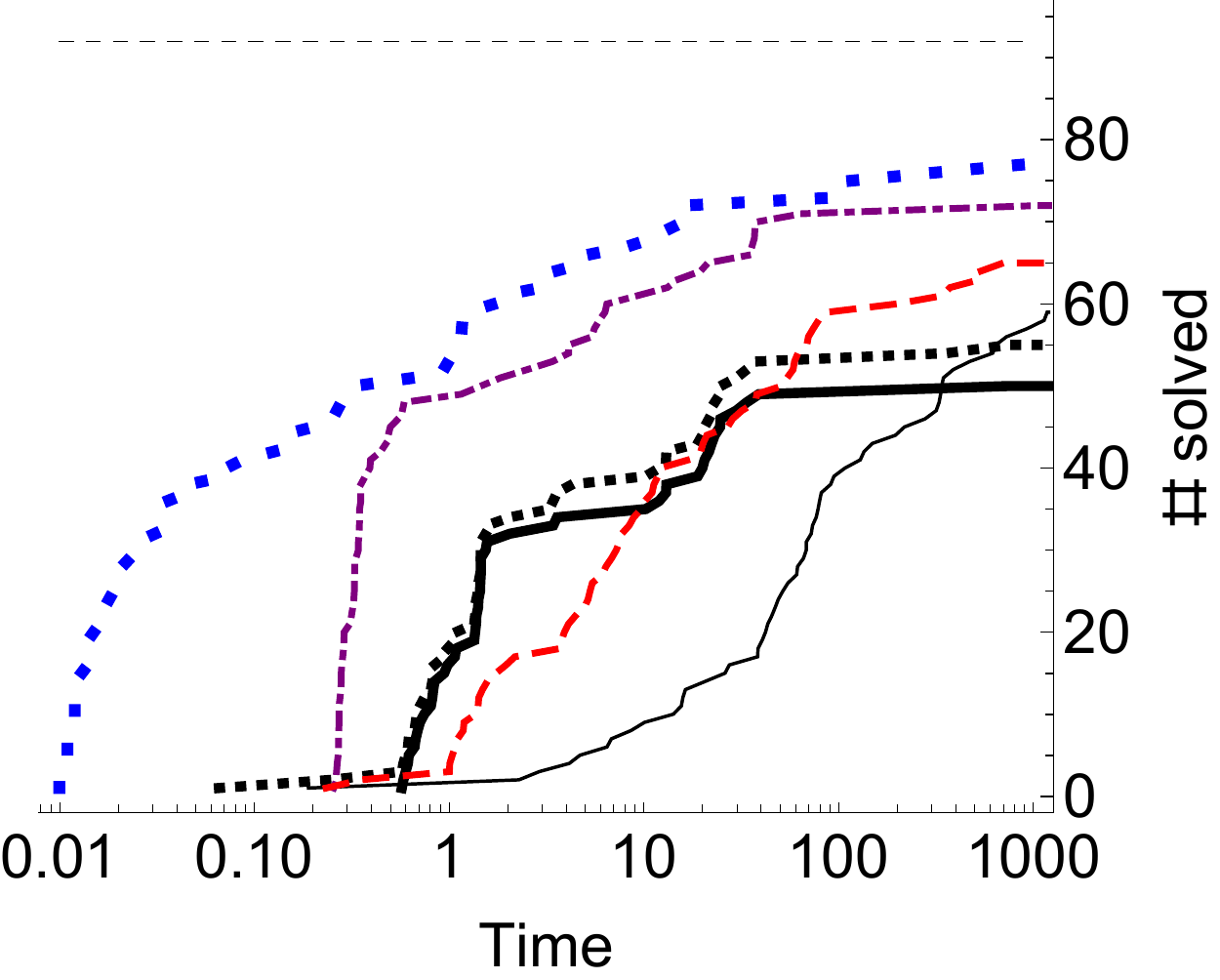} 

            \hspace*{1.5em}
            \includegraphics[height=.2\textheight]{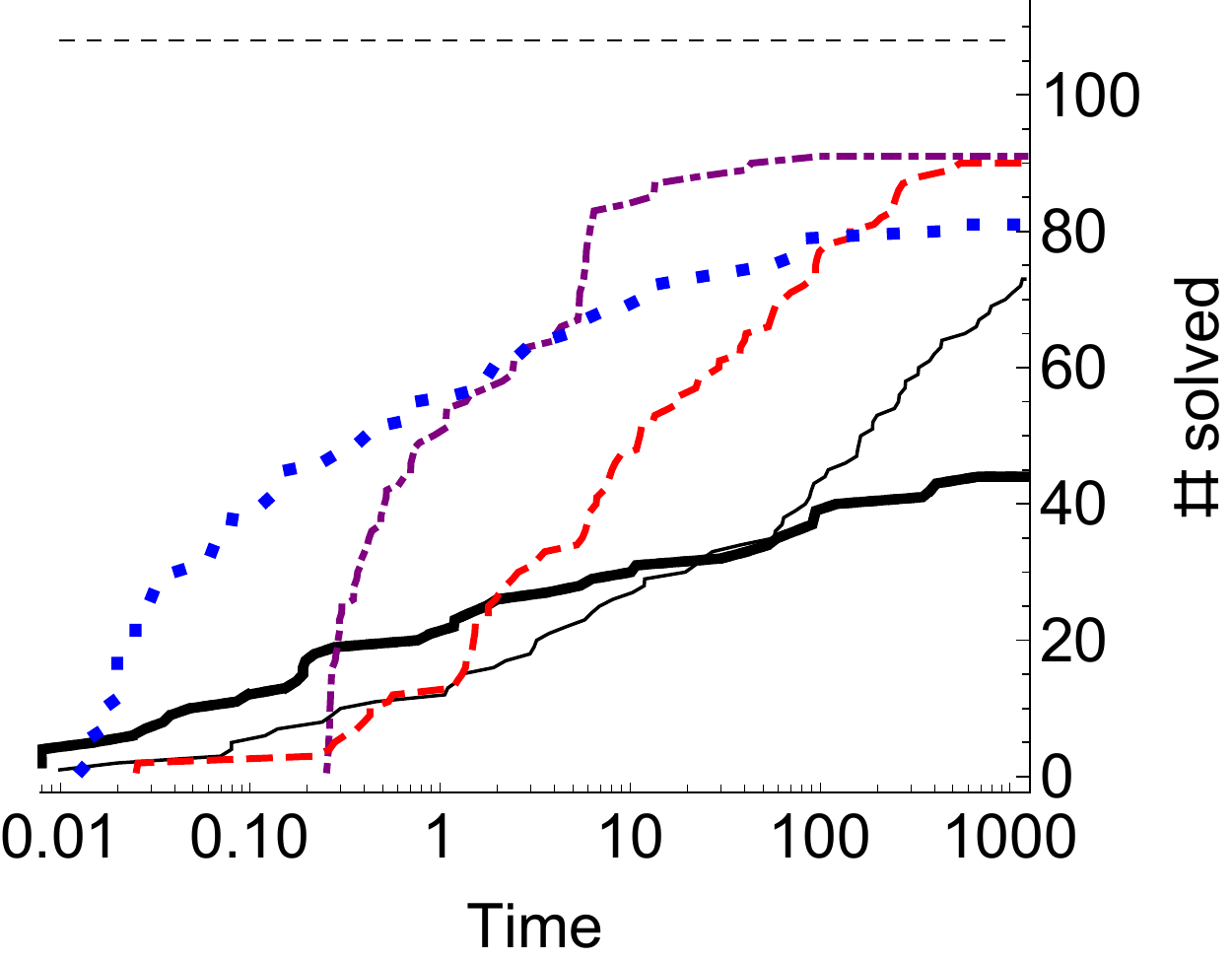} 
        \end{minipage}}
    %
    \caption{\label{f:bench} Results on benchmark problems ($\Unsat$ (upper) and $\Sat$ (lower) instances).}
\end{figure} 

The second experiment is based on the following two sets of benchmark problems.

\begin{itemize}
\item \emph{Linear arithmetic (LA) benchmark.}
We translated instances in the \verb|QF_LRA| section of the SMT-LIB benchmarks into FPA instances by simply converting data sorts (from $\mathbb{R}$ to $\mathbb{F}_{11,53}$), operators, etc.
Each real constant is converted to an exact FP constant if possible, otherwise they are converted to a rounded value.
Since the set is large, we picked the instances whose originals were solvable by Z3 within the 30s.\footnote{Instances using the \texttt{ite} function were omitted; many of \texttt{LassoRanker} and \texttt{meti-tarski} instances were removed to balance the number.}
Because the proposed method abstracts the rounding modes of FPA operators, 
we parameterized the rounding modes in formulas and represented them by unconstrained variables. 
Also, every free and unassigned variable was asserted that it is not $\NaN$.

\item \emph{Griggio benchmark.}
The Griggio suite, taken from the SMT-LIB benchmarks,\footnote{\url{http://smtlib.cs.uiowa.edu/benchmarks.shtml}.} offers challenging problems for bit-blasting solvers and has been used in several experiments~\cite{Marre2017,Brain2019,Zitoun2020}.
Here, we use the suite as a standard FPA problem set to evaluate our solver.
Some instances involve multiple precisions and concrete rounding modes are given in most cases.
\end{itemize}

Fig.~\ref{f:bench} shows the cactus plots of the number of solved instances versus time (with semi-logarithmic scale), assuming each instance is solved in parallel.
For our solver, results counting $\Unknown$s are also shown with dotted lines.
The setting 3 (using \textsc{CVC4}) or 2 (incremental) solved more instances than the others for LA or Griggio, respectively.
Detailed statistics are described in Appendix~\ref{s:xp:bench:stat}.
For instances for which the solution was previously unknown, the solution obtained by any solvers was assumed to be correct;
two instances of LA were excluded because the outputs did not match among the solvers.

\subsection{Discussions}

Regarding \textbf{RQ1}, we obtained results that were better or comparable to those of other solvers, except for the Griggio benchmark. 
Our RIA-based method (with the appropriate settings) solved the most $\Unsat$ and $\Sat$ instances for the BMC and LA sets (in which the rounding modes are not specified).
The results for Griggio, a benchmark that includes instances designed for dedicated solvers, on the other hand, were dismal.
We believe that our method is inefficient for instances where the decision depends on rounding mode settings or combinations of normal and special FP numbers.
Overall, our method was able to compete with other dedicated FPA solvers.
Also, no solver performed outstandingly well in all the experiments.
For example, \textsc{CVC4} and \textsc{MathSAT} performed well for some BMC instances, but they resulted in a lot of timeouts and out-of-memory errors.
In LA, only the RIA solver could solve 21 instances.


As for \textbf{RQ2}, $\Unknown$ results were less than $10\%$ for most of the problem sets, whereas around $30\%$ were $\Unknown$s for $\Unsat$ instances of LA.
From the results, we consider that the impact of $\Unknown$s were rather small because the imact of execution time on scalability was much greater (cf. the result of BMC and the fact that many of the instances could not be solved within 1200s).
The main cause was that we inhibit falsifying inequalities $\f \Rel_-^\neg \g$ of the weak extension when $\f$ or $\g$ can be $\NaN$.
This cause can be dealt with by case analyses, e.g., detection of assignments to a free variable, and we have actually implemented some analyses in our translators.
Otherwise, $\Unknown$s occur more often as the number of operations increases and by the \emph{wrapping effect} (cf. the rotation system in BMC).
Reduction of errors using e.g. Affine form instead of interval vectors will be a future work.
In addition, we use rounding operators in Def.~\ref{d:rounding} based on a mild estimation of errors.
We consider that the use of linear formulas improved the efficiency of the solving process while providing sufficient accuracy.


The RIA incremental solver performed better than the non-incremental solver using \textsc{Z3} for LA and Griggio;
for LA, non-incremental solving using \textsc{CVC4} was better than incremental probably due to the performance of \textsc{CVC4} in solving linear formulas.
In BMC, non-incremental performed better than incremental because all the instances required double precision.
When it is decidable with a coarser precision and/or the lemmas learned along the way accelerate the solving process, the incremental solver outperforms.

%
%
%

%


%

\section{Conclusion}

We have proposed an IA logic to approximate FPA formulas and a dedicated solver using RA solvers of \textsc{CVC4} and \textsc{Z3}.
Despite using an off-the-shelf RA solver, we obtained experimental results that were competitive with those of other FPA solvers;
we confirmed that our solver is effective for a subset of FPA (BMC and LA) where rounding modes are parameterized.
In the experiments, although the solver was shown inefficient for the FPA benchmark Griggio,
it solved the most numbers of instances for two such problem sets.


\bibliographystyle{splncs03}
\bibliography{delta}

\newpage

\appendix

\section{Formal Verification using Why3}
\label{s:why3}

We have (partially) verified the correctness of the proposed method using \textsc{Why3},\footnote{\url{http://why3.lri.fr}} a verification platform with plugged-in theorem provers.
Lemmas~\ref{lm:sound} and \ref{lm:ineq} have been verified as follows.
We first defined a real interval type and a predicate ``$x \in \x$'' (where $x\in\mathbb{R}^*$ and $\x\in\mathbb{I}^*$) in \textsc{Why3}'s input language.
Then, Lemma~\ref{lm:sound} was verified for the four operators $+$,$-$,$\times$, and $\div$.
For every operator $\circ$, we implemented the interval extension as procedure $\f: \mathbb{I}^*\times\mathbb{I}^*\to\mathbb{I}^*$
and verified the Hoare triple $\{\x,\y\in\mathbb{I}^*\}\ \r := \f(\x, \y)\ \{\forall m \in \mathbb{M}, x\in\x \land y\in\y \Rightarrow \circ(m, x, y) \in \r\}$.
It resulted in a number of verification conditions and they were discharged using the back-end provers i.e. \textsc{Alt-Ergo}\footnote{\url{https://alt-ergo.ocamlpro.com/}} and \textsc{Coq}.\footnote{\url{https://coq.inria.fr/}}
Next, Lemma~\ref{lm:ineq} was verified for the predicates $\x \Rel_?^{[\neg]} [0]$ and $\x \Rel_?^{[\neg]} \y$ where $\x$ and $\y$ are limited to identifiers.
We defined the comparison operators as \textsc{Why3} predicates and their properties (cf. Lemma~\ref{lm:ineq}) as \textsc{Why3} lemmas.
The lemmas were then proved using \textsc{Alt-Ergo} and \textsc{Coq}.
The \textsc{Why3} description is available at \url{https://github.com/dsksh/fp_rint_why3}.

\section{Interval Comparison Operators}
\label{s:compop}

We denote an interval expression $\f \setminus \{\NaN\}$ by $\f_{\setminus\NaN}$.
Let $\backsim$ be the \emph{accurate subtraction} operator;
$\f \backsim \g$ is interpreted as
$[\LB{f}_{\setminus\NaN}-\UB{g}_{\setminus\NaN}, {\UB{f}_{\setminus\NaN}-\LB{g}_{\setminus\NaN}}]$ appended with $\{\NaN\}$ if $\f$ or $\g$ contains $\NaN$.
The comparison operators in IA are defined in a logic on $\mathbb{R}^+$ as follows:
\begin{align*}
    \f \RelP_?^{[\neg]} \g &~:\Leftrightarrow~ \f \backsim \g \RelP_?^{[\neg]} [0], & \\
    \f \RelP_- [0]        &~:\Leftrightarrow~ \UB{f}_{\setminus\NaN} \RelP 0, & 
    \f \RelP_+ [0]        &~:\Leftrightarrow~ \NaN \not\in \f \land \LB{f} \RelP 0, \\
    \f \RelP_-^{\neg} [0] &~:\Leftrightarrow \NaN\!\in\!\f \lor \LB{f}_{\setminus\NaN}\!\RelP^\neg\!0, &
    \f \RelP_+^\neg [0]   &~:\Leftrightarrow~ \UB{f}_{\setminus\NaN} \RelP^\neg 0, \\[.5em]
    \f =_- \g           &~:\Leftrightarrow~ \f \geq_- \g \land \g \geq_- \f, & 
    \f =_+ \g           &~:\Leftrightarrow~ \NaN \not\in \f,\g \land \LB{f} \!=\! \UB{f} \!=\! \LB{g} \!=\! \UB{g}, \\
    \f \neq_- \g        &~:\Leftrightarrow~ \f <_- \g \lor \g <_- \f, & 
    \f \neq_+ \g        &~:\Leftrightarrow~ \f <_+ \g \lor \g <_+ \f, \\[.5em]
    \f \equiv_- \g      &~:\Leftrightarrow~ \rlap{$(\NaN\in\f \LAnd \NaN\in\g) \lor \f =_- \g,$} & \\
    \f \not\equiv_- [0] &~:\Leftrightarrow~ \f \neq_- \g, &
    \f \equiv_+ \g      &~:\Leftrightarrow~ \f =_+ \g, \\
    \f \not\equiv_+ \g  &~:\Leftrightarrow~ \rlap{$(\NaN \not\in \f \lor \NaN \not\in \g) \land \f \neq_+ \g,$} 
\end{align*}
where $\RelP \in \{\geq, >\}$ and $\RelP^\neg \in \{<, \leq\}$.

\section{Incremental Solving Process}
\label{s:impl:inc}

The solver script described in Sect.~\ref{s:impl:solver} runs two sub-process that follows Alg.~\ref{a:inc} in parallel.
It assumes FPA with a \emph{precision bound} $(\mathit{eb},\mathit{sb})$ that represents the finest precision assumed in $\phi$.
$\textsc{Encode}(?,\phi)$ generates an interval extension $\mathbm{\phi}^?$ with the abstract precision mode, which encodes while leaving the precision parameters (e.g. $\mathit{ed}$ and $\mathit{em}$) undefined.
%
The main loop of Alg.~\ref{a:inc} tries to solve under several precisions configured from coarser to exact ones.
Note that, the bounds for the maximum normal FP numbers is not modified throughout the process to attain the soundness.
The \textsc{CheckSatAssuming} process invokes the RA solver of Z3~\cite{Jovanovi2012}, which combines linear programming (LP) and cylindrical algebraic decomposition (CAD) techniques within the CDCL framework.

\begin{algorithm}[tb]
    \SetAlgoLined
    \SetKwInOut{Input}{Input}\SetKwInOut{Output}{Output}
    \Input{Precision bound $(\mathit{eb},\mathit{sb})$, $?\in\{-,+\}$, FPA formula $\phi$}
    \Output{$\Unsat$, $\Sat$ or $\Unknown$}
    \BlankLine
    $\mathbm{\phi}^?$ := $\textsc{Encode}(?, \phi)$\;
    \For{$(\mathit{eb}',\mathit{sb}') :\in [(4,4); (5,11); (8,24); (11,53); (15,113)])$}{
        $\mathit{eb}'' := \min \{\mathit{eb},\mathit{eb}'\}$; $\mathit{sb}'' := \min \{\mathit{sb},\mathit{sb}'\}$\;
        $r$ := $\textsc{CheckSatAssuming}(\mathbm{\phi}^?, \textsc{DefConstants}(\mathit{eb}'',\mathit{sb}''))$\;
        \textbf{if}\ {$(?=- \land r=\Unsat) \lor (?=+ \land r=\Sat)$}\ \textbf{then return} $r$ \textbf{end}
    }
    \Return{$\Unknown$}\;
    \caption{Incremental solving process.}
    \label{a:inc}
\end{algorithm}

\section{Target Systems of BMC}
\label{s:xp:bmc:spec}

Three systems are specified by the following recursive functions:
\begin{align*}
    &\text{Integrator:~} &
    y(i) &:= x(i) + 0.9 y(i\!-\!1), \\
    &\text{Filter:~} &
    y(i) &:= 
    \begin{pmatrix}
        c_1 x(i) - c_2 y_1(i\!-\!1) - c_3 y_2(i\!-\!1) \\
        y_1(i\!-\!1)
    \end{pmatrix}, \\
    &\text{Rotation:~} & 
    y(i) &:= 
    \begin{pmatrix}
        c_4 & -c_5 \\
        c_5 & c_4
    \end{pmatrix} {y}(i\!-\!1),
\end{align*}
where 
the parameters are 
$c_1 := 0.058167$, 
$c_2 := 1.4891$, 
$c_3 := 0.88367$, 
$c_4 := 0.86602540303$ and
$c_5 := 0.5$, and
the input $x(i)$ is constrained as $x(i) \in [-1,1]$.
We assume the initial condition $y(0) := 0$.
A bounded path of output values is represented by a sequence $y(1) \cdots y(k)$.
The output value $y(k)$ or $y_1(k)$ was compared with the threshold $\mathit{th}$ in the experiment.

\section{Statistics on Benchmark Problems}
\label{s:xp:bench:stat}

Tables~\ref{t:r:la}--\ref{t:r:griggio} show a breakdown of the results by problem set and solver.
The first few columns show the number of $\Unsat$/$\Sat$ results and their percentage of all instances, followed by the number of $\Unknown$ results.
``To'' and ``Oom'' represent the numbers of runs that resulted in timeout and out of memory, respectively.
The last two columns show the total time taken to have all the conclusive results and the maximum memory usage during the (conclusive) solving processes.
Table~\ref{t:r:la} additionally shows the numbers or erroneous runs (``Err.'').
For the LA instances, \textsc{Colibri} often terminated due to an error apparently caused by the use of unbounded rounding mode variables.


\begin{table}[h!]
    \centering
    \caption{\label{t:r:la} Results for the LA benchmark ($46$ $\Unsat$ and $107$ $\Sat$ instances; there were also $70$ $\Unknown$ instances not shown in the table).}
    \begin{tabular}{l|rrrrrrrrrr} \hline
        Solver & Unsat &~ Sat &\hspace{1.5em} Rate solved &~ Unk. &~ To & Oom & Err. & Total time & Max Mem. \\
        \hline
        \hline
        \textsc{RIA} (1) & $29$       & $77$       & $63.0\%\!+\!72.0\%$ & $31$ & $83$ & $1$ & $0$ & $17500$s & $168$MB \\
        \textsc{RIA} (2) & $29$       & $36$       & $63.0\%\!+\!33.6\%$ & $9$  & $144$ & $3$ & $0$ & $933$s & $303$MB \\
        \textsc{RIA} (3) & ${\bf 30}$ & ${\bf 98}$ & ${\bf 65.2}\%\!+\!{\bf 91.6}\%$ & $22$ & $54$ & $17$ & $0$ & $13700$s & -- \\
        \textsc{Colibri} & $19$       & $50$       & $41.3\%\!+\!46.7\%$ & $16$ & $71$ & $5$ & $61$ & $677$s & -- \\
        \textsc{CVC4}    & ${\bf 30}$ & $52$       & ${\bf 65.2}\%\!+\!48.6\%$ & $0$  & $17$ & $122$ & $0$ & $10800$s & -- \\
        \textsc{MathSAT} & ${\bf 30}$ & $9$        & ${\bf 65.2}\%\!+~8.4\%$ & $0$  & $83$ & $99$ & $0$ & $478$s & $120$MB \\
        \textsc{Z3}      & ${\bf 30}$ & $81$       & ${\bf 65.2}\%\!+\!75.7\%$ & $0$ & $16$ & $94$ & $0$ & $9530$s & $597$MB \\
        \hline
    \end{tabular}
\vspace{3em}
    \centering
    \caption{\label{t:r:griggio} Results for the Griggio benchmark ($92$ $\Unsat$, $108$ $\Sat$ and $14$ $\Unknown$ instances).
      No errors occurred, and all the results were sound.}
    \begin{tabular}{l|rrrrrrrr} \hline
        Solver & Unsat &~ Sat &\hspace{1.5em} Rate solved &~ Unk. &~ To &~ Oom & Total time & Max mem. \\
        \hline
        \hline
        \textsc{RIA} (1) & $34$       & $39$       & $37.0\%\!+\!36.1\%$ & $3$ & $138$ & $0$ & $3700$s & $38$MB \\
        \textsc{RIA} (2) & $50$       & $44$       & $54.3\%\!+\!40.7\%$ & $5$ & $115$ & $0$ & $4010$s & $55$MB \\
        \textsc{RIA} (3) & $42$       & $37$       & $45.7\%\!+\!34.3\%$ & $2$ & $124$ & $9$ & $8980$s & -- \\
        \textsc{Colibri} & $72$       & ${\bf 91}$ & $78.3\%\!+\!{\bf 84.3}\%$ & $2$ & $49$ & $0$ & $1880$s & -- \\
        \textsc{CVC4}    & $65$       & $90$       & $70.7\%\!+\!83.3\%$ & $0$ & $1$ & $0$ & $8690$s & -- \\
        \textsc{MathSAT} & ${\bf 77}$ & $81$       & ${\bf 83.7}\%\!+\!75.0\%$ & $0$  & $56$ & $0$ & $3110$s & $56$MB \\
        \textsc{Z3}      & $73$       & $59$       & $64.1\%\!+\!67.6\%$ & $0$ & $66$ & $0$ & $24500$s & $1060$MB \\
        \hline
    \end{tabular}
\end{table}

Tables~\ref{t:ncomp:la}--\ref{t:ncomp:griggio} show the number of instances solvable with each solver but another solver did not solve.

\newcommand{\Sp}{\phantom{1}}
\begin{table}[h!]
    \centering
    \caption{\label{t:ncomp:la} Numbers of $\Unsat+\Sat$ instances from the LA set the solver in the row solved but the solver in the column could not.}
    \begin{tabular}{l|@{\hskip .5em}r@{\hskip 2em}r@{\hskip 2em}r@{\hskip 2em}r@{\hskip 1em}r@{\hskip 1em}r@{\hskip 0em}r@{\hskip 3em}r} \hline
        Solver &~ \textsc{RIA}(2) & \textsc{RIA}(3) & \textsc{Colibri} & \textsc{CVC4} & \textsc{MathSAT} & \textsc{Z3} \\
        \hline
        \textsc{RIA} (2) & --\Sp\Sp\Sp       & $\Sp 0$ + $0$ & $16$ + $14$    & $\Sp 3$ + $\Sp 7$ & $10$    + $28$ & $\Sp 3$ + $\Sp 0$ \\
        \textsc{RIA} (3) & $\Sp 1$ + $62$    & --\Sp\Sp\Sp   & $17$ + $51$    & $\Sp 4$ + $52$    & $11$    + $90$ & $\Sp 4$ + $25$ \\
        \textsc{Colibri} & $\Sp 7$ + $28$    & $\Sp 7$ + $3$ & --\Sp\Sp\Sp    & $\Sp 3$ + $11$    & $\Sp 6$ + $47$ & $\Sp 3$ + $\Sp 5$ \\
        \textsc{CVC4}    & $\Sp 4$ + $25$    & $\Sp 4$ + $8$ & $13$ + $15$    & --\Sp\Sp\Sp       & $13$    + $45$ & $\Sp 0$ + $\Sp 2$ \\
        \textsc{MathSAT} & $12$    + $\Sp 1$ & $12$    + $1$ & $17$ + $\Sp 6$ & $14$    + $\Sp 0$ & --\Sp\Sp\Sp    & $14$    + $\Sp 0$ \\
        \textsc{Z3}      & $\Sp 4$ + $46$    & $\Sp 4$ + $9$ & $13$ + $37$    & $\Sp 0$ + $30$    & $13$    + $73$ & --\Sp\Sp\Sp \\
        \hline
    \end{tabular}
\vspace{5em}
    \centering
    \caption{\label{t:ncomp:griggio} Numbers of $\Unsat+\Sat$ instances from the Griggio set the solver in the row solved but the solver in the column could not.}
    \begin{tabular}{l|@{\hskip .5em}r@{\hskip 2em}r@{\hskip 2em}r@{\hskip 2em}r@{\hskip 1em}r@{\hskip 1em}r@{\hskip 0em}r@{\hskip 3em}r} \hline
        Solver &~ \textsc{RIA}(2) & \textsc{RIA}(3) & \textsc{Colibri} & \textsc{CVC4} & \textsc{MathSAT} & \textsc{Z3} \\
        \hline
        \textsc{RIA} (2) & --\Sp\Sp\Sp    & $\Sp 9$ + $21$ & $\Sp 3$ + $\Sp 4$ & $\Sp 2$ + $\Sp 0$ & $11$    + $\Sp 1$ & $\Sp 2$ + $\Sp 3$ \\
        \textsc{RIA} (3) & $\Sp 1$ + $14$ & --\Sp\Sp\Sp    & $\Sp 1$ + $\Sp 5$ & $\Sp 0$ + $\Sp 4$ & $\Sp 9$ + $\Sp 5$ & $\Sp 0$ + $\Sp 1$ \\
        \textsc{Colibri} & $25$    + $51$ & $31$    + $59$ & --\Sp\Sp\Sp       & $12$    + $16$    & $14$    + $17$    & $16$    + $30$ \\
        \textsc{CVC4}    & $17$    + $46$ & $23$    + $57$ & $\Sp 5$ + $15$    & --\Sp\Sp\Sp       & $13$    + $18$    & $\Sp 6$ + $24$ \\
        \textsc{MathSAT} & $38$    + $38$ & $44$    + $49$ & $19$    + $\Sp 7$ & $25$    + $\Sp 9$ & --\Sp\Sp\Sp       & $31$    + $17$ \\
        \textsc{Z3}      & $11$    + $32$ & $17$    + $37$ & $\Sp 3$ + $12$    & $\Sp 0$ + $\Sp 7$ & $13$    + $\Sp 9$ & --\Sp\Sp\Sp \\
        \hline
    \end{tabular}
\end{table}

\end{document}